\documentclass[a4paper,10pt,titlepage]{elsart}
\usepackage[latin1]{inputenc}

\usepackage[dvips]{graphicx}
\usepackage{amsfonts} 
\usepackage{mathrsfs}
\usepackage{amsmath}
\usepackage{longtable}
\usepackage{amssymb}
\usepackage{color}
\usepackage{epsfig}

\def\RR{{ I\!\!R}}
\def\arctan{\hbox{Arctan}}
\def\cotan{\hbox{cotan}}
\def\cotanh{\hbox{cotanh}}
\def\d{\displaystyle}

\def\ff{{\mathscr F}}

\def\IH{IH}
\graphicspath{{figures/}{figures/intro/}{figures/correction/}{figures/resultats/}{figures/annexe/}}

\begin{document}
\clearpage
\setcounter{page}{1}

\begin{frontmatter}
\title{Analysis and optimisation of the tuning of the  twelfths for a clarinet resonator}
\author[marseille]{V. Debut},
\corauth[cor]{Corresponding author}
\ead{debut@lma.cnrs-mrs.fr}
\author[marseille]{J. Kergomard},
\author[paris]{F. Lalo\"{e}}
\address[marseille]{Laboratoire de M\'ecanique et d'Acoustique-CNRS UPR 7051, 31 chemin Joseph Aiguier, 13402 Marseille cedex 20, France}
\address[paris]{Laboratoire Kastler-Brossel, Dept de physique de l'ENS UMR CNRS 8552, 24 rue Lhomond, 75231 Paris cedex 05, France}
\begin{abstract}
Even if the tuning between the first and second register of a
clarinet has been optimized by instrument makers, the lowest
twelfths remain slightly too large (inharmonicity). In this
article, we study the problem from two different points of view.
First, we systematically review various physical reasons why this
inharmonicity may take place, and the effect of different bore
perturbations inserted in cylindrical instruments such as bore
flare, open and closed holes, taper, temperature gradients, visco-thermal effects, etc. Applications to a real clarinet resonator and
comparisons with impedance measurements are then presented. 
A commonly accepted idea is that the register hole is the dominant cause for this inharmonicity: it is natural to expect that opening this hole will shift the position of the resonances of the instrument to higher frequencies, except of course for the note for which the hole is exactly at the pressure node. We show that the real situation is actually more complicated because other effects, such as open holes or bore taper and bell, introduce resonance shifts that are comparable but with opposite sign, so that a relatively good overall compensation takes place. This is checked by experimental and theoretical studies of the acoustical impedance curves of a clarinet. The origin of the observed inharmonicity in playing frequencies is therefore different, maybe related to the reed or the vocal tract. In a second part, we use an elementary model of the clarinet in order
to isolate the effect of the register hole: a perfect cylindrical
tube without closed holes. Optimization techniques are then used
to calculate an optimum location for the register hole (without
taking into account the use of the register hole as a B flat tone
hole); the result turns out to be close to the location chosen by
clarinet makers. Finally, attempts are made numerically to improve
the situation by introducing small perturbations in the higher
part of the cylindrical resonator, but no satisfactory improvement
is obtained.
\end{abstract}


\end{frontmatter}

\section{Introduction}
Many contributions, either experimental or theoretical, have been
made in the past to improve our understanding of single reed
woodwind instruments (see
e.g.\cite{Nederveen:69,McIntyre:83,Gilbert:86,Dalmontbis:00}). Our
knowledge of the linear behaviour of the resonator is now
satisfactory, which probably explains why most of the recent
literature deals with the understanding of the sound production
and oscillation regimes, an inherently non-linear problem. This
does not mean that all interesting questions concerning the linear
behaviour of the resonator have been solved. For instance, Benade
\cite{Hoekje:95} proposed in the seventies some basic ideas and
methods allowing the qualities of a wind instrument to be
characterised (impedance peak alignment), but more detailed
explorations along this line would be useful; Meynial and
Kergomard \cite{Meynial:88} proposed relatively recently simple
acoustical systems that shift the scale of a woodwind by a given
micro-interval. Concerning the relation between linear properties
and non-linear oscillations, ref.\cite{Dalmont:95} discusses in
general how it is possible to predict the emission frequencies,
and even some aspects of the clean intonation and tone colour,
which naturally leads to the design of modified instruments as
soon as appropriate optimization criteria are defined. Recently,
optimization techniques have been used in order to define
longitudinal profiles of brass instruments \cite{Kausel:01}.

\paragraph*{}An example of an unsolved problem is given by the clarinet. While many musicians agree that
it is now a well tuned instrument, the twelfths corresponding to the four lowest tones remain slightly too large, by 20 or 30 cents \cite{Rendall:71,fobes:web,Deplus:private}. One could imagine many reasons for this problem:
deviation of the bore from purely cylindrical shape, existence of
open or closed side holes (cavities), temperature gradient, etc. A
generally accepted idea states nevertheless that the main reason
resides in the register hole, because it cannot be ideally placed for every fingering.
\paragraph*{}The first purpose of the present article is to analyze the origin of this tuning
problem in detail. We will actually see that the occurrence of
wide twelfths at either end of the clarinet register is not an
obvious problem in terms of linear impedance theory, in
contradiction with common wisdom. In real clarinets, various
effects of opposite signs tend to compensate each other, at least
partially (role of the flaring bell for instance). The second
purpose of the article is to investigate whether or not it is
possible to design a register hole allowing the first two complete 
registers to be perfectly well tuned (the definition of the
registers will be given in the next section), ignoring all the
other sources for inharmonicity except the register hole. We will then reason on a
simplified shape, cylindrical and without tone holes, the
different values of the tones being adjusted simply by choosing
different lengths for the tube. Two questions will be discussed :

\begin{description}
\item i) what is the optimum location of the register hole for the
tuning of the twelfths intervals between the first two registers?
\item ii) is it possible to propose some simple system, located
upstream from the highest tone hole, to provide a good correction
to the residual tuning defaults?
\end{description}
The outline of this paper is as follows: the second section recalls some general features of the
clarinet. The third section is devoted to the first study: it analyzes the length corrections and their variation with frequency for the main kinds of discontinuities encountered in wind instruments. It ends with an application to a real clarinet resonator in order to find whether the register hole is the main culprit of the clarinet tuning defect; calculations are compared with measurements. A validation of the calculations to first order in the ratio of the length correction to the wavelength is also given as well as a complete analysis of the effects of an open and closed tone holes. Then, sections 4 and 5 are concerned with the second part of our study and answer questions i) and ii) respectively. Some useful formulae are
given in appendix \ref{anx:un}.
\section{Generalities on the  clarinet}

\subsection{Definitions}
\begin{figure}[htpb]
\centering
\includegraphics[width=12cm]{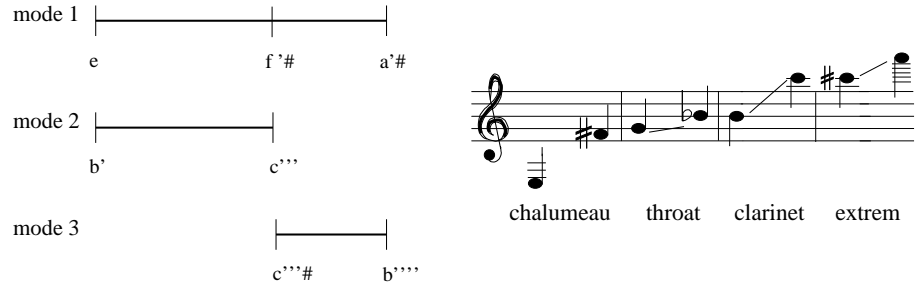}
\caption{\textit{Compass and operative resonance frequency for a clarinet.}}
\label{fig:compass}
\end{figure}

\begin{figure}[htpb]
\centering
\includegraphics[width=12cm]{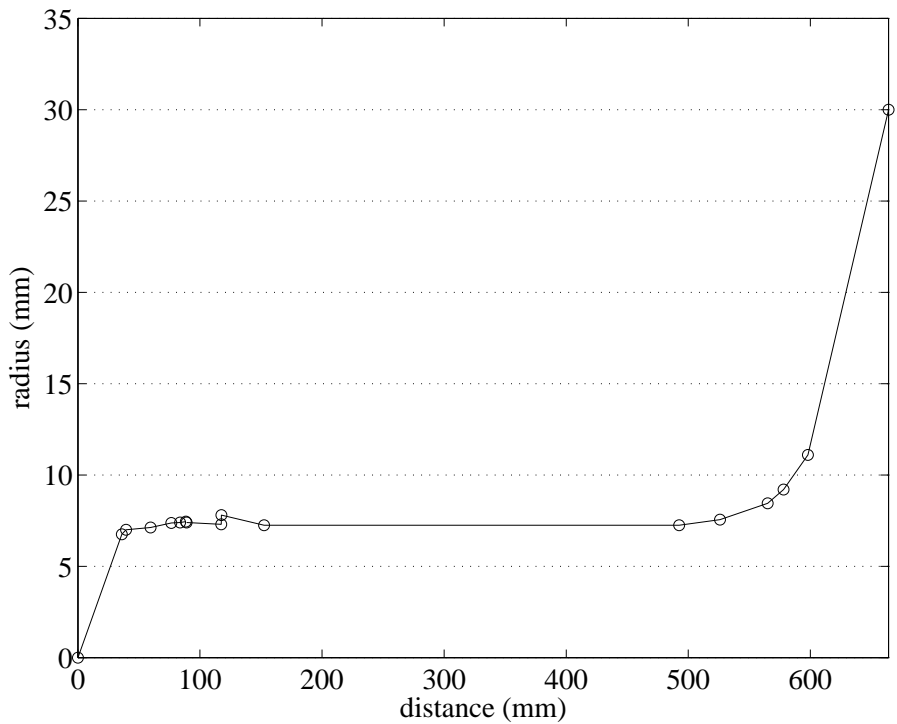}
\caption{\textit{Radius of a clarinet as a function of the distance from the reed tip (see table \ref{tab:data}).}}
\label{fig:profil}
\end{figure}

For a given fingering, several resonance frequencies of
the resonator occur, and several oscillation regimes can be obtained. The term register is used to describe the set of tones obtained for the same
regime: the \textit{first register} involves the tones corresponding to the
first mode (i.e. the first resonance frequency) of the resonator, the
frequencies being denoted $f_{1\text{ }}$, and the \textit{second register }
involves the tones corresponding to the second mode, the frequencies being denoted
$f_{2}$ (where $f_2$ is approximately $3f_{1}$). The opening of the register hole allows the musician to "jump" from a given tone of the
first register to the corresponding tone of the second register, one octave
and a fifth above, the first two registers covering ideally more than 3
octaves\footnote{We notice that while this division into two registers seems
obvious to the scientist, the analysis of the characteristics of the tone
colour is not so clear, and musicians divide this musical compass in four
registers:  chalumeau, throat, clarinet and extrem \cite{Rendall:71}.}. For the fingerings  corresponding to the higher tones of the first register, i.e. for fingerings from \textit{f$\,^\prime$\#} to \textit{a$\,^\prime$\#}\footnote{The notation  system adopted in this paper is the same as that used by Baines \cite{Baines:77}.}, the opening of the register hole is not sufficient in
order to get the proper twelfth, and it is necessary to modify slightly the
fingering in order to ensure the good tuning. Thus, in order to play the corresponding twelfths (i.e tones from \textit{c$\,^{\prime\prime\prime}$\#}), the third resonance, corresponding to an interval of a 17th, is used, corresponding to that the physicist can call the third register (see figure \ref{fig:compass}).
\paragraph*{}The most commonly used clarinet is the B-flat clarinet which sounds one tone
below the written note. Five parts can be distinguished on a clarinet: the mouthpiece, the barrel, the upper and lower joints and the bell. To first approximation, the clarinet is a cylindrical instrument, but a detailed determination of the geometrical dimensions of the instrument reveals deviations from this regular, cylindrical shape (see figure \ref{fig:profil}) over parts which are short with respect to the wavelength. Moreover, the existence of tone and register holes, bore irregularities, thermal gradient and dispersion due to visco-thermal effects alter the resonance frequencies and their harmonicity; they must be included for a detailed study of the tuning of the instrument.
\subsection{Inharmonicity and sound production}
The sound of the clarinet is produced by self-sustained
oscillations. Theory shows \cite{Grand:96} that, at low sound
intensity, the playing frequencies are determined by the zero of
the imaginary part of the input admittance of the resonator (they
are therefore very close to the frequencies of the maxima of the
input impedance modulus). In addition, the reed has also a small
influence on the playing frequency, for two different reasons: the
volume velocity created by the motion of the reed movement, which
adds to that produced by the pressure difference across the reed
opening ; the frequency pulling effect of the resonance of the
reed (at higher frequencies) combined with its damping. It can be
shown that these two effects can be taken into account by
introducing appropriate length corrections to the instrument,
which are almost independent of the frequency
\cite{Nederveen:69,Dalmont:95}. At higher sound intensities, the
playing frequency also depends on other impedance peaks; it can
 slightly change under the effect of the inharmonicity of
the resonances of the resonator. A proper harmonicity of the first
two resonances is therefore important (even for a single note);
then, the sound frequency remains independent of the playing
level, even if the first two resonances have similar magnitude,
and good "impedance peak cooperation" \cite{Benade:76_ch20} ensures a
stable sound emission. This question is actually intricate because
it involves the influence of many parameters. Here, for the sake
of simplicity we limit ourselve to low intensity sounds; in other
words, our goal will just be to achieve a perfect tuning at least
for pianissimo levels.\newline
In this article, we will distinguish between \textit{inter-resonance} inharmonicity and \textit{inter-register} inharmonicity. The former
refers to the inharmonicity between the first two resonance
frequencies of the resonator for a given fingering; the latter
defines the ratio between the resonance frequencies of the first
impedance peak when the register hole is closed and the second one when the register hole is
open \footnote{The targeted interval is  a
pure twelfth, i.e. almost a ratio of 3. Actually for a tempered
scale, the ratio is slightly different, because the tempered
intervals are different from the natural ones: the exact value is
$2^{19/12}=2.9966\;$. The relative difference is 0.11\%, i.e. 2
cents (it is the difference between the tempered fifth and the
harmonic one, called the ''skhisma''). Nevertheless this
difference is very small and for simplicity in what follows we
will ignore it (2 cents are almost inaudible, and within the
''tunability range'', especially for higher notes
\cite{Leipp:76}).}

\subsection{Method and approximation}
The shape of a clarinet is very close to a pure cylinder. For a given fingering of the chalumeau, with first
resonance frequency $f_{1}$, an effective length $\ell _{eff}$ can be
defined as
\begin{equation*}
\ell _{eff}=c/4f_{1}\;,
\end{equation*}
where $c$ is the speed of sound. A first approximation of this length is given by the physical
length between the input of the instrument (taking into account a length
correction for the above mentionned reed effects) and the first open tone
hole position, where the tube is assumed to be cut. All the perturbations from
the cylindrical shape, such as enlargments or contractions, closed tone
holes, as well as the precise effects of the open tone hole (depending on
the downstream geometry and radiation property), can be regarded as small
corrections to this first approximation leading to the correct value, $\ell_{eff}$. For the lowest tone, for which all holes are closed, an equivalent length of the flaring horn is used as in \cite{Nederveen:69}. If the resonator was purely cylindrical and lossless, the second resonance frequency
would be exactly $f_2=3f_{1}$ . However all the perturbations produce corrections
slightly different from those on the first resonance: this creates the inter-resonance inharmonicity (for
inter-register inharmonicity, the opening of the register hole also needs to be
taken into account). Here, we are actually concerned not by
the effect itself of the perturbations from the cylindrical shape, but by the
variation of this effect between the two registers. \newline
The concept of frequency dependent length corrections is an adequate tool for this study. As in reference \cite{Meynial:88} the corrections upstream of the first open hole can be
calculated separately for each kind of perturbation, with respect to the
effective input of the tube (taking into account the reed effects) where the imaginary part of the admittance must be zero for self-sustained oscillations. The
frequency dependence of the length corrections for all fingerings provides the inharmonicity. Only the effect of the part that is upstream the first open hole needs to be studied for each
fingering.

\paragraph*{}All calculations are carried out by ignoring the different kinds of dissipation (due
to visco-thermal effects in the boundary layers, to radiation, etc...):
weak dissipation is known to have a negligible effect on the resonance
frequencies \cite{Kergomard:81}. In Appendix B the effect of the resistance of a small hole is
discussed, and even if nonlinear effects are taken into account, it is shown
to be negligible. A consequence is the systematic use of purely imaginary
impedances, which means for the resonance condition, that the input impedance is infinite. Perturbation to the planar mode theory is classically taken into
account using lumped elements representing the effects of higher order,
evanescent modes of the tubes.
\section{Length correction and inharmonicity produced by a small perturbation: analysis of the different effects for a real clarinet resonator}
In this section  we study the analytic expressions of length corrections and inharmonicity
of the first two resonance frequencies associated with small perturbations to a purely cylindrical shape. This formulation gives a good idea of the inharmonicity encountered on a real clarinet resonator. 

\subsection{Length corrections: definition}
The effect of a geometrical perturbation  on a cylindrical resonator may be expressed conveniently in terms of a length correction to the main tube, denoted $\Delta \ell $. Using an exact
formulation of $\Delta \ell $ is possible but in this study, we use an approximation of length corrections to first order in the ratio of the length correction to the wavelength: this gives a sufficiently accurate determination of the resonance frequencies. Moreover, the length corrections associated with different perturbations can be simply added as we will see in a particular example.
The perturbation is located at a distance $\ell$ from the effective input, which is the origin of the coordinates once the clarinet embouchure has been replaced by an equivalent cylinder and the reed effects have been taken into account (see \ref{subsub:calculations}).

\subsection{Inharmonicity of the resonance frequencies}
Inharmonicity can be defined as the  relative difference between the
resonance frequency  $f_{n}$ and $n$ times the first resonance $f_{1}$, as
follows:
\begin{equation}
\IH=\frac{f_{n}-nf_{1}}{nf_{1}}=\frac{\ell _{eff}+\Delta \ell _{1}}{\ell
_{eff}+\Delta \ell _{n}}-1=-\frac{\Delta \ell _{n}-\Delta \ell _{1}}{\ell
_{eff}}+o(\frac{\Delta \ell }{\ell _{eff}})\;,  \label{eq:inharmonicite}
\end{equation}
where $\ell _{eff}$ is the acoustic length of the unperturbed system  and $\Delta \ell _{n}$ is the length correction associated to the $nth$ resonance.
\newline
In the case $\IH>0$  the basic intervals are enlarged; when $\IH<0$  the intervals are reduced. In the present paper, results are
given in \textit{cent}; the cent is the micro-interval equal to one hundredth of
a tempered semi-tone:
\begin{equation*}
1\;cent=5.78\;10^{-4}\mbox{\quad and \quad}\IH)_{cents}={\frac{\IH}{%
5.78\;10^{-4}}}\;.
\end{equation*}
We notice that the smallest frequency deviation perceptible by the human ear, estimated to be 4 cents, corresponds to $\IH\simeq 0.25\%$.

\subsection{Length corrections and inharmonicity formulae for acoustic basic systems}\label{par:discontinuites} 
This section contains analysis of inharmonicity associated with the simple acoustical systems usually encountered in a clarinet, and depicted in table \ref{tab:deltaL}. From the calculation of first order length corrections\footnote{The expression \textit{first order length corrections} refers to length corrections expanded to first order in the ratio of the length correction to the wavelength.}, analytical expressions of inharmonicity between the first and second resonance frequencies are derived. For exact formulations of length corrections, the reader is referred to appendix \ref{anx:un}.
\begin{table}[htbp]
\begin{center}
\begin{tabular}{|c|c|}
\hline
\raisebox{1.2cm}{register hole} & \includegraphics[width=10cm]{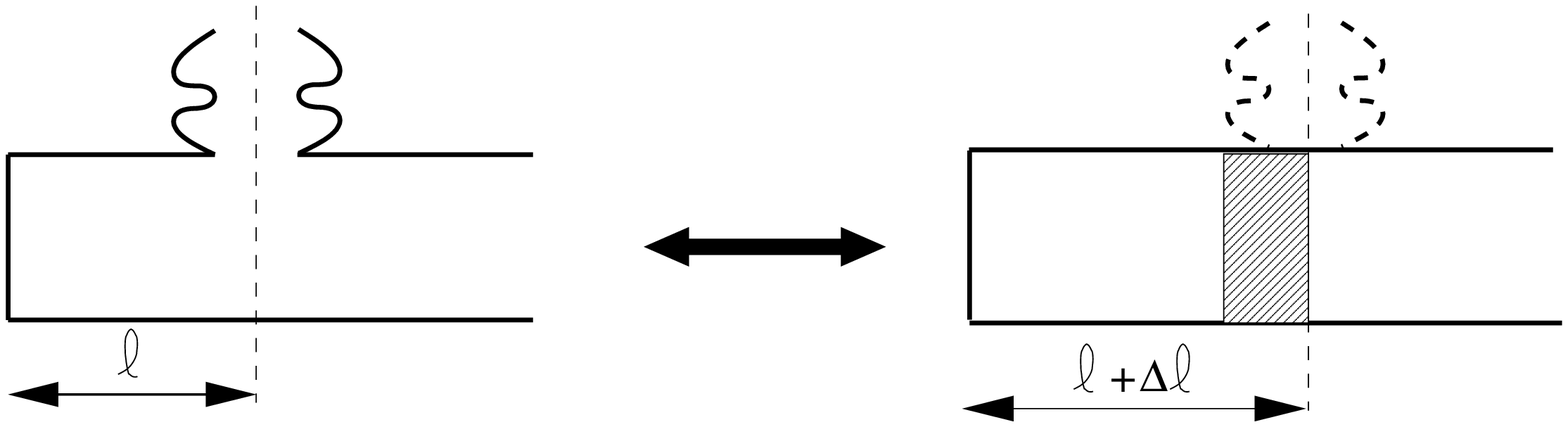}
\\
\hline
\raisebox{1.2cm}{tone hole} & \includegraphics[width=10cm]{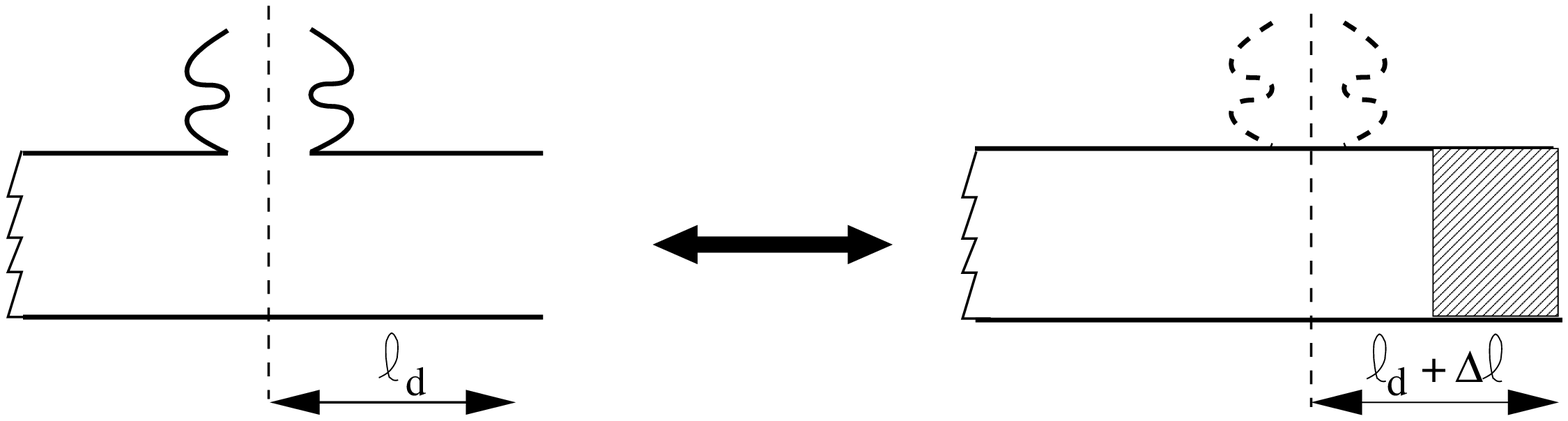}
\\ \hline
\raisebox{1.2cm}{closed-hole} & %
\includegraphics[width=10cm]{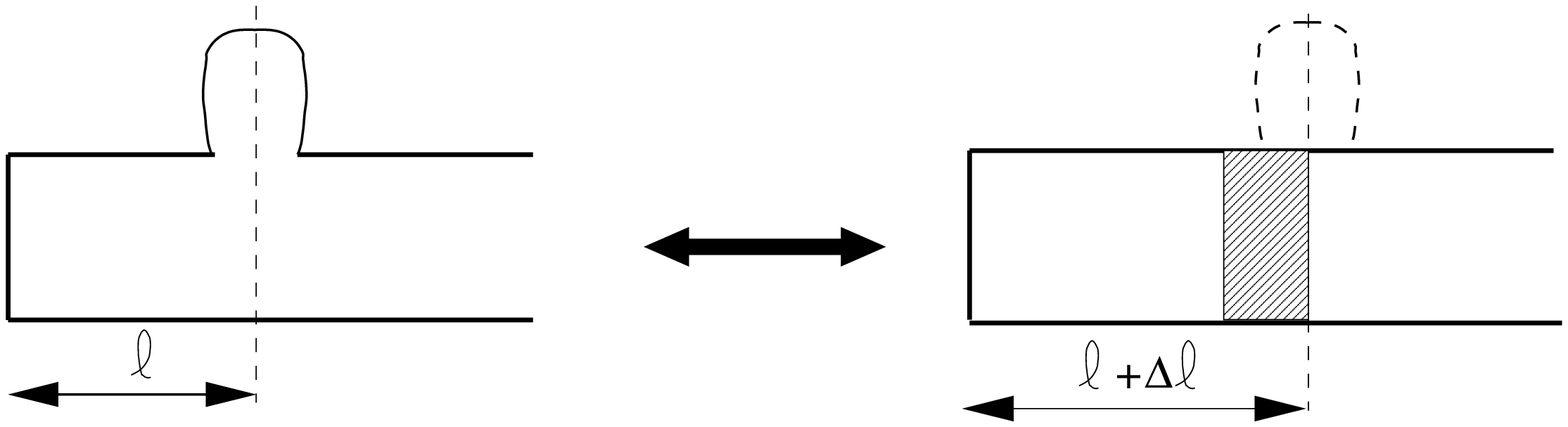} \\ \hline
\raisebox{1.0cm}{abrupt change in cross section } &  \includegraphics[width=10cm]{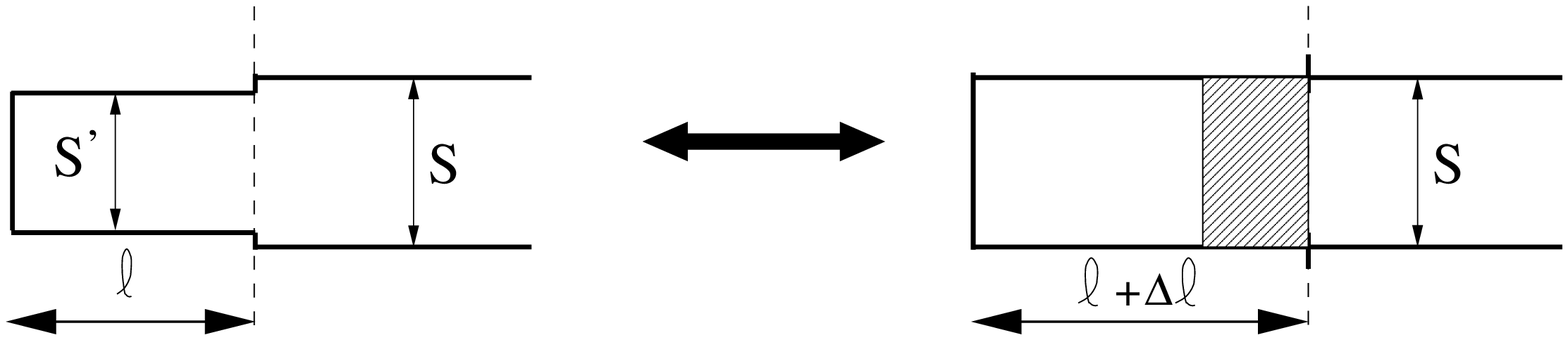} \\ 
\raisebox{.4cm}{in the upper part of the instrument} & \\ \hline
\raisebox{1.0cm}{localized enlargement/contraction} & %
\includegraphics[width=10cm]{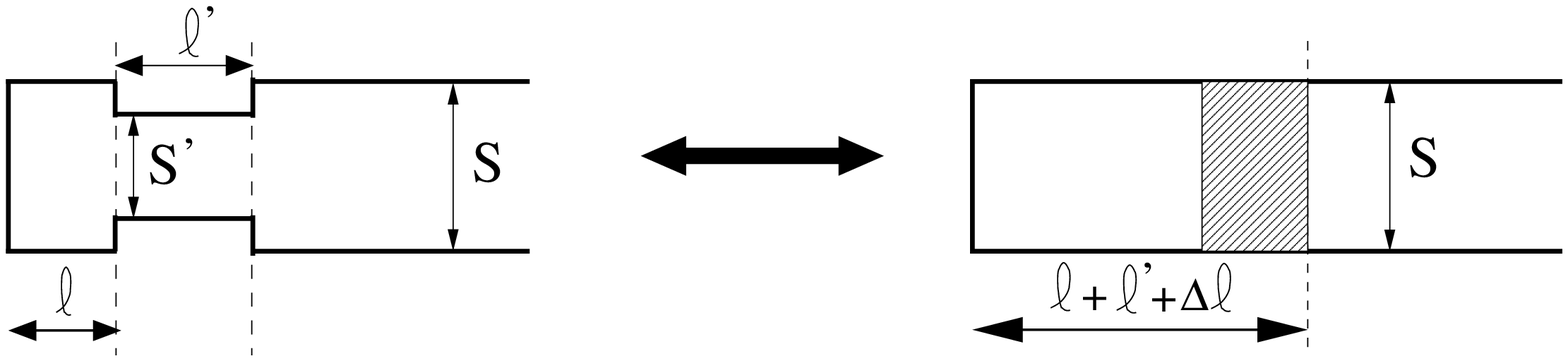} \\ 
\raisebox{.4cm}{in the upper part of the instrument} & \\ \hline
\raisebox{1.2cm}{truncated cone} & 
\includegraphics[width=10cm]{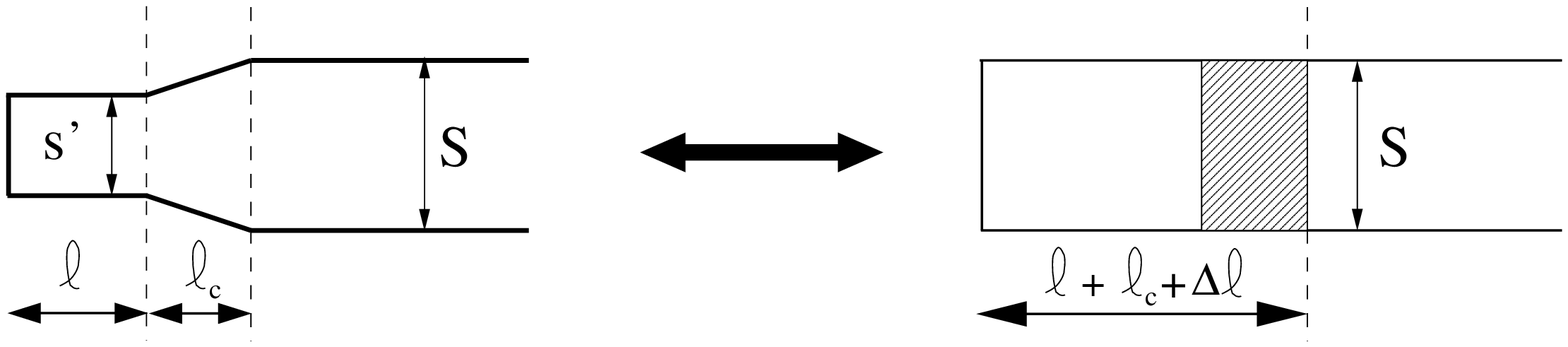} \\ 
\raisebox{.4cm}{in the upper part of the instrument} & \\ 
\hline
\end{tabular}
\end{center}
\caption{\textit{Acoustic basic systems and associated length corrections.}}
\label{tab:deltaL}
\end{table}

\subsubsection{Insertion of a side hole} \label{subsub:sidehole}
This section uses results obtained by Meynial and Kergomard \cite{Meynial:88} concerning the effects of the insertion of an admittance branched in parallel and an impedance branched in series, on a cylindrical tube. Starting from the tone hole model proposed by Dubos \textit{et al.}  in \cite{Dubos:99}, we propose a new calculation of an equivalent length that includes the effects of both the series and shunt impedances. Consequences for the cases of a register hole, an open side hole and a closed hole are subsequently discussed. 
\paragraph*{General formulation for an admittance in parallel and an impedance in series}
We consider a side hole, or any discontinuity inserted in parallel,
branched on a straight cylindrical tube (a typical example is the register hole shown in table \ref{tab:deltaL}). $Y$ is the admittance of the inserted discontinuity \footnote{Throughout the paper the admittance is defined as a ratio of an acoustic volume velocity to an acoustic pressure.} and $Y_{up}$ and $Y_{down}$ are the main tube admittances upstream and downstream the discontinuity respectively. The following equation can be written at the location of the discontinuity ($x=\ell $), 
\begin{equation}
Y_{up}=Y+Y_{down}\;,  
\label{eq:admittance}
\end{equation}
\noindent Looking backwards to the top-end of the resonator and writing \- $Y_{down}=-jY_{c}\tan k(\ell +\Delta \ell )\;$, we get from equation (\ref{eq:admittance}):
\begin{equation}
-jY_{c}\tan k\ell =Y-jY_{c}\tan k(\ell +\Delta \ell )\;,
\label{eq:admittancebis}
\end{equation}
where $\Delta \ell $ is the length correction, $Y_{c}=S/\rho c$ is the characteristic admittance of the main tube ($\rho $ is the density of air and $S$ the cross section area of the main tube), $k=2\pi f/c$ is the wavenumber and $j=\sqrt{-1}\;$. After some algebra, the following result is obtained:
\begin{equation}
k\Delta \ell =\arctan \left( \frac{-j\frac{Y}{Y_{c}}\cos ^{2}k\ell }{1-\frac{%
1}{2}j\frac{Y}{Y_{c}}\sin 2k\ell }\right) \quad ,  \label{eq:deltaL_general}
\end{equation}
which can be approximated in the limit of small $Y/Y_{c}$ by:
\begin{equation}
k\Delta \ell \simeq -j\frac{Y}{Y_{c}}\cos ^{2}k\ell \quad .
\label{eq:deltaL_parallele}
\end{equation}
For the case of a series impedance $Z$ branched on a cylindrical tube, similar expressions are found to be, respectively:
\begin{equation}
k\Delta \ell =\arctan \left( \frac{-j\frac{Z}{Z_{c}}\sin ^{2}k\ell }{1+\frac{%
1}{2}j\frac{Z}{Z_{c}}\sin 2k\ell }\right) \quad ,  \label{eq:deltaL_general_serie}
\end{equation}
and 
\begin{equation}
k\Delta \ell \simeq -j\frac{Z}{Z_{c}}\sin ^{2}k\ell \quad .
\label{eq:deltaL_serie}
\end{equation}
Formulae (\ref{eq:deltaL_parallele}) and (\ref{eq:deltaL_serie}) can also be deduced from Rayleigh's variational principle \cite{Rayleigh:45}. They are useful for different kinds of small discontinuities (see reference \cite{Meynial:88}), but for some particular cases, other formulae need to be derived as we see below.

\paragraph*{Side hole}
Referring to \cite{Dubos:99} for the tone hole model, the acoustic pressure and the acoustic volume velocity at both sides of the hole can be related by the following transfer matrix:
\begin{equation}
\left(
\begin{array}{cc}
A & B \\
C & D 
\end{array}
\right)
=\frac{1}{1-Y_sZ_a/4}
\left(
\begin{array}{cc}
1+Y_sZ_a/4 & Z_a \\
Y_s & 1+Y_sZ_a/4 
\end{array}
\right)\;,
\label{eq:transfer_matrix}
\end{equation}
where $Z_a$ and $Y_s$ are the series impedance and shunt admittance respectively. This formulation emphasizes the dual role of the two terms $Z_a$ and $Y_s$, as opposed to formulations based on the classical T or $\Pi$ equivalent circuits. Manipulation of equation (\ref{eq:transfer_matrix}) gives the length correction:
\begin{equation}
-j\,Y_c\,\tan k(\ell+\Delta \ell)=\frac{-Y_s+DY_{\ell}}{A-Z_aY_{\ell}}\;,
\end{equation}
where $Y_{\ell}=-j\,Y_c\,\tan k\ell\;.$ Finally:
\begin{equation}
k\Delta\ell=\arctan\left (\frac{-jY_s/Y_c\cos^2 k\ell-jZ_a/Z_c \sin^2 k\ell}{1+Y_sZ_a/4+j(Z_a/Z_c-Y_s/Y_c)\sin k\ell\cos k\ell}\right )\;.
\label{eq:total_tonehole}
\end{equation}
As a result, equation (\ref{eq:total_tonehole}) is a general expression for the length correction of a tube branched off a cylindrical tube. It gives the influence of both the series and shunt impedances in the determination of the equivalent length. No assumption has been made concerning the dimensions and shape of the hole: it is therefore possible to derive first order approximations for special cases as tone hole, register hole or closed hole from the relative influence of $Z_a$ and $Y_s$. For instance, in the case of an open hole, both $Z_a$ and $Y_s$ are inductive. With the low frequencies approximation, equation (\ref{eq:total_tonehole}) shows that $\Delta \ell$ is mostly determined by the shunt admittance. If $Y_s/Y_c\ll1$, the length correction is small and equation (\ref{eq:total_tonehole}) gives equation (\ref{eq:deltaL_parallele}); this is the case of the register hole.

\begin{figure}[hbtp]
\centering
\includegraphics[width=12cm]{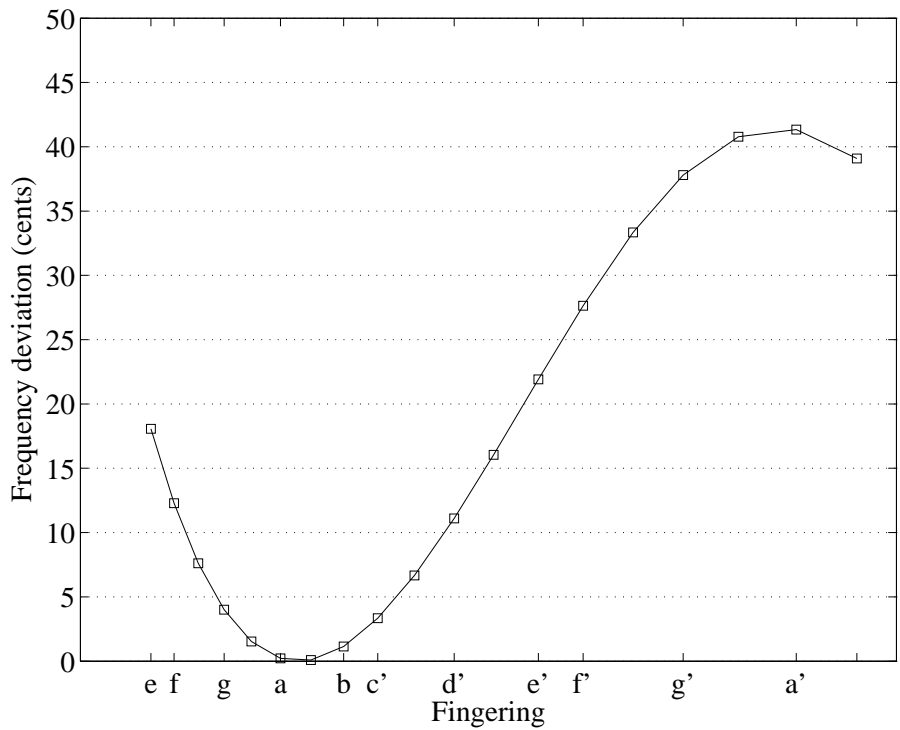}
\caption{\textit{Shift of the second resonance frequency due to a
register hole of radius $r=1.4\;mm$, height $h=13\;mm$, located a distance $\ell =140\;mm$ from the effective input. For the b flat fingering, the register hole does not alter the second vibration mode  since, in this case, the register hole is located at one third the effective length.}}
\label{fig:inharmo_trou_registre}
\end{figure}

\begin{figure}[hbtp]
\centering
\includegraphics[width=6cm]{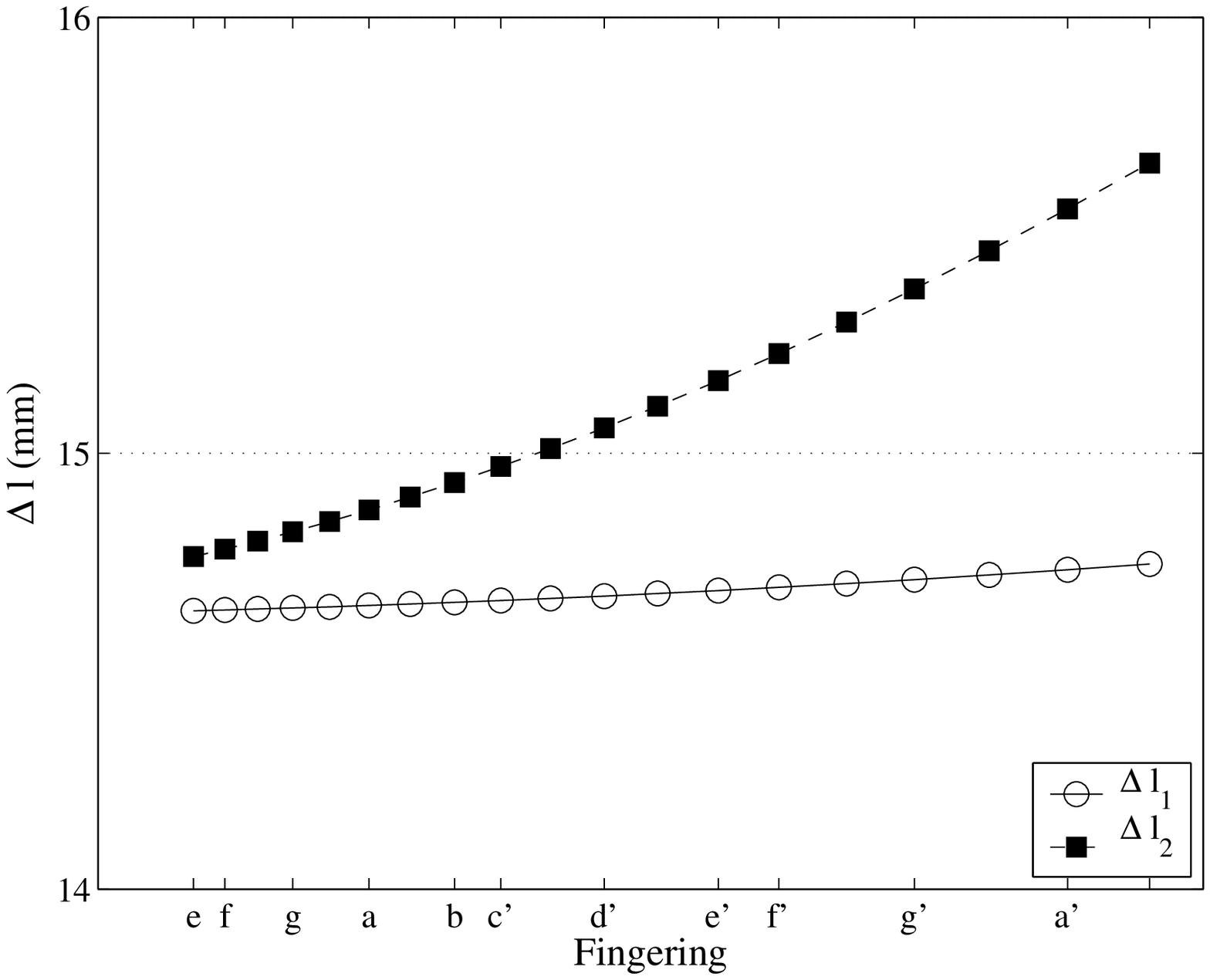} 
\includegraphics[width=6cm]{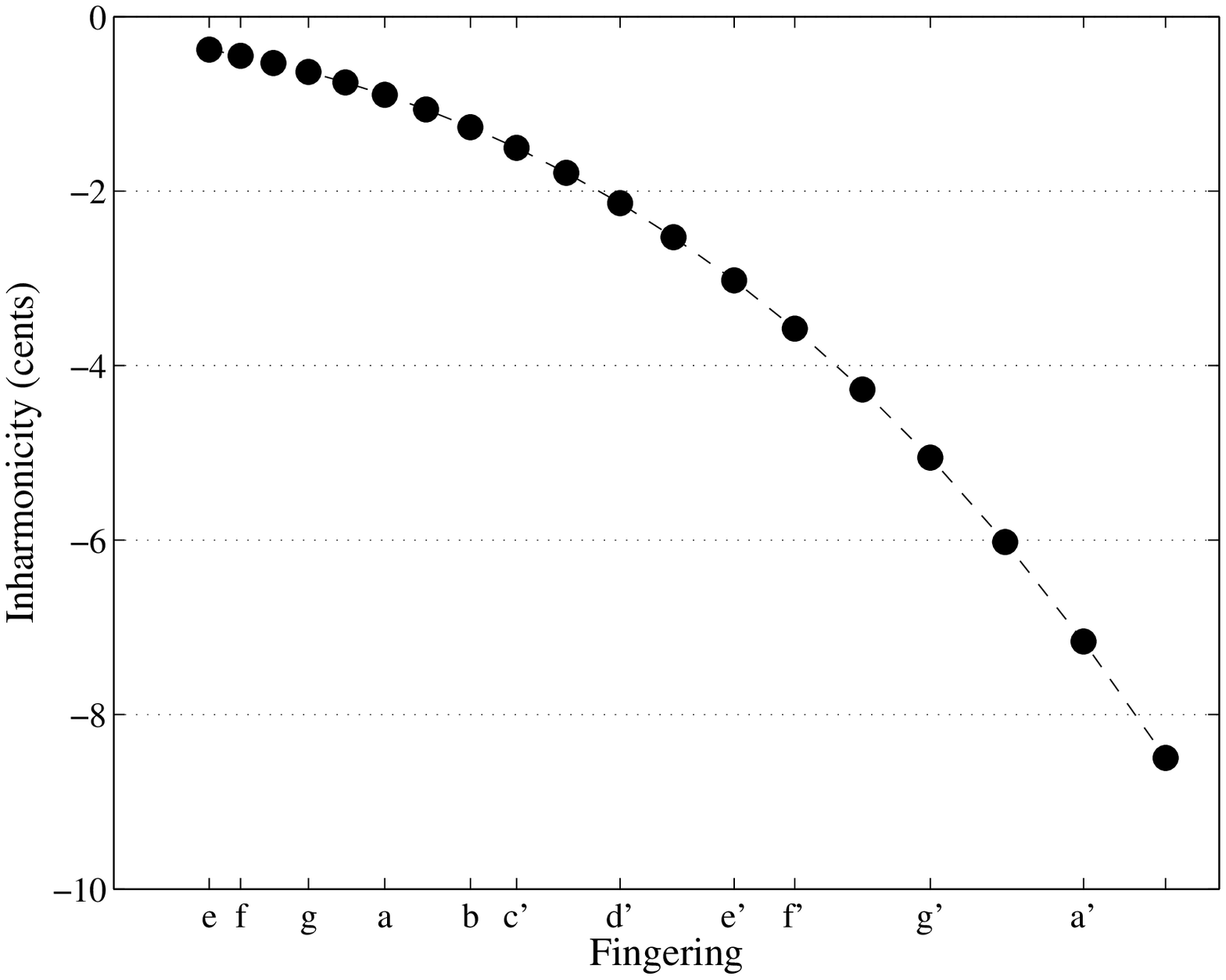}
\caption{\textit{Length corrections (left) and corresponding inter-resonance inharmonicity
(right) for a tone hole of radius $r=4.5\;mm$, height $h=4\;mm$, located at a distance $\ell _{d}=30\;mm$ from the open end for each tube length.}}
\label{fig:inharmo_gros_trou}
\end{figure}

\paragraph*{Open hole: register hole}
The register hole is a side chimney with small diameter but large height located a distance $\ell\simeq140\;mm$ down the effective input (typically $r=1.5\;mm$, $h=13\;mm$). 
Using \cite{Dubos:99} to calculate the elements of the model, we obtain the series impedance as:
\begin{equation}
Z_a=j\omega\rho \ell_a^{(o)} /S\;,
\label{eq:Zao}
\end{equation}
where $\ell_a^{(o)}$ is the series length correction for an open hole. The shunt impedance is:
\begin{equation}
Z_s=Y_s^{-1}=j\frac{\rho c}{S_h} \left(\tan [k(h+h_m+h_r)+jkh_s]\right) \;,
\end{equation}
where $S_h$ is the cross-sectional area of the hole, and $h_m$, $h_s$ and $h_r$ are height corrections associated with the matching volume, the higher order modes impedance and radiation, respectively. Expressions for these lengths can be found in \cite{Dubos:99,Dalmont:01} (see also \cite{Dalmont:02} for a correction to \cite{Dubos:99} for $h_s$) and are given hereafter:
\begin{eqnarray}
&& \ell_a^{(o)}=-r\delta^2[1.78\tanh (1.84(h+h_m)/r)+0.940+0.540\delta+0.285\delta^2]^{-1}\\
&& h_m=\frac{r\delta}{8}(1+0.207\delta^3)\\
&& h_s=r(0.82-0.193\delta -1.09\delta^2+1.27\delta^3-0.71\delta^4)\\
&& h_r=0.82r\left(\frac{1+(0.77kr)^2}{1+0.77kr}\right)^{-1}
\end{eqnarray}
where $\delta=r/R$, $k$ is the wavenumber of the played tone and the subscript $o$ is used to refer to the open hole case. Since the low frequency limit is valid, the shunt impedance is well approximated by: \begin{equation}
Z_s=j\frac{\rho c}{S_h} kh^{\prime}\;,
\end{equation} where:
\begin{equation}
h^{\prime}=h+h_m+h_s+h_r\;.
\label{eq:h_prime}
\end{equation}
This equation allows us to discuss the relative influence of the series and shunt impedances. Calculation gives $\mid Y_sZ_a\mid\simeq 0.0045 \%\;$ which confirms that the series impedance can be ignored in the case of a long chimney with a small diameter. Finally,  equations (\ref{eq:total_tonehole}) is identical to (\ref{eq:deltaL_parallele}), which provides:
\begin{equation}
k\Delta \ell \simeq -{\frac{S_h\ell }{Sh^{\prime }}}\frac{\cos ^{2}k\ell }{k\ell }\;,  
\label{eq:deltaL}
\end{equation}
where $\ell$ is the distance of the hole from the effective input. Equation (\ref{eq:deltaL}) shows that a small open side hole always introduces a negative length correction, i.e. an increase in the
resonance frequencies. It also shows that $\Delta \ell $ decreases as the inverse of the square of
frequency. Therefore, the register hole opening affects the first resonance frequencies much more than the second one, which is an important requirement for the clarinet to overblow correctly \cite{Benade:76}.\newline
\noindent From equation (\ref{eq:deltaL}), since the register hole has no effect on tones of the first register for which it is closed, the frequency shift of the second resonance frequency due to the register hole opening can be derived. We then get:
\begin{equation}
\IH \simeq -\frac{\Delta \ell_2}{\ell_{eff}} \simeq \frac{2}{3\pi }\frac{S_h}{Sh^{\prime }}\frac{cos^{2}k_2\ell }{k_2}\;,
\label{eq:inharmo_reg}
\end{equation}
where $k_2=3k_{1}$, $k_1$ being the wavenumber of the fundamental frequency of the played tone. This expression shows that the register hole opening generates positive inharmonicity by pulling the second vibration mode upward in frequency at both ends of the register scale and has no effect at a pressure node as it is well known (see figure \ref{fig:inharmo_trou_registre} where $r=1.4\;mm$, $h=13\;mm$ and $\ell=140\;mm$). It also appears that the two geometrical parameters which control inter-resonance inharmonicity are $Sh^{\prime }/S_h$ and $\ell$.
\paragraph*{Open hole: tone hole} \label{par:tonehole}
In the limit of zero frequency, the shunt admittance of an open hole increases to infinity (see equation (\ref{eq:deltaL_parallele})) and the main tube behaves as if it was cut at the location of the hole. The ratio $Y_s/Y_{c}$ becomes large for tone holes and equation (\ref{eq:total_tonehole}) can no longer be used since the perturbation cannot be considered as small anymore. As a consequence, $\Delta \ell$ is no longer seen as an extension of the entire air column of length $\ell_{eff}$ but as an additional length to $\ell$ the distance from the effective input to the hole location. For an open hole of height $h^{\prime }$ (see eq.(\ref{eq:h_prime})), radius $r$ at distance $\ell _{d}$ from the open end, the length
correction is now obtained from:
\begin{equation}
-j\,Y_c\,\cotan\,k\Delta \ell =\frac{Y_s+DY_{\ell}}{A+Z_aY_{\ell}}\;,
\label{eq:gros_trou}
\end{equation}
where now $Y_{\ell}=-j\;Y_c\;\cot k\ell_d\;.$ Thus: 
\begin{equation}
k\Delta\ell=\arctan\left(\frac{-jY_c[(1+Z_aY_s/4)-jZ_aY_c\cot k\ell_d]}{Y_s-(1+Z_aY_s/4)jY_c\cot k\ell_d}\right)\;.
\label{eq:gros_tonehole}
\end{equation}
For a real clarinet using the expressions for $Z_a$, $Y_s$ and $Y_c$ for an open hole, we see that the quantity $\mid Z_aY_s/4\mid$ is small compared to unity, except for the four bottom tones for which the order of magnitude is to 0.02 . In the same way, the term in bracket at the numerator is close to unity. This leads us to assume that $Z_a$ has a negligeable effect in this study for open holes especially since we focus on inharmonicity. Setting the term in bracket to unity in equation (\ref{eq:gros_tonehole}) and using Taylor's formula to the
third order in $k\ell _{d}$ and to the first order in $k\Delta \ell $, the tone hole equivalent length becomes: 
\begin{equation}
\Delta \ell \simeq \frac{1}{q+{\frac{1}{\ell_{d}(1+(k\ell _{d})^{2}/3)}}}\;,
\label{eq:deltaL_troudenote}
\end{equation}
where $q=r^{2}/(R^{2}h^{\prime })$ is relative to the geometry of the tone
hole. From equation (\ref{eq:deltaL_troudenote}),the length correction becomes a constant value in the lower frequency limit given by $\Delta \ell =\ell _{d}\ell _{hole}/(\ell _{d}+\ell _{hole})\;,$where $\ell _{hole}=h^{\prime }\,S/S_h$. When frequency increases, the tone hole length correction increases too. Then, any upper resonance frequency relative to the lowest one is flattened, and negative inharmonicity is generated as it can be seen in figure \ref{fig:inharmo_gros_trou} (for a tone hole geometry given by $r=4.5\;mm$, $h=4\;mm$ located, for each length of tube, at a distance $\ell_d=30\;mm$ from the open end).

\paragraph*{}Concerning the inter-resonance inharmonicity, equation (\ref{eq:deltaL_troudenote}) leads to the expression
\begin{equation}
\IH\simeq -\frac{16}{3\pi }\frac{k_{1}^{3}\ell _{d}^{3}}{(1+q\ell _{d})^{2}}\;.
\label{eq:IH_trou_ouvert}
\end{equation}
Discrepancy between equation (\ref{eq:IH_trou_ouvert}) and numerical calculations from equation (\ref{eq:gros_tonehole}) gives less than a 4-cent difference for all fingerings except for three tones, $c^{\prime}\#$, $f^{\prime}$ and $g^{\prime}$ for which results differ of an amount to about 8 cents. This  appears to be due to the use of the Taylor expansion of the\textit{ $\cotan$} function to obtain equation (\ref{eq:deltaL_troudenote}). Finally, equation (\ref{eq:IH_trou_ouvert}) is a simple analytic formula which also appears reliable, and will be used in the next. It also gives an interesting feature: the freedom it gives the designer to control inharmonicity due to open hole in the choice of the hole size. Large tone holes would be prefered to reduce inharmonicity of the first two resonance frequencies.

\begin{figure}[hbtp]
\centering
\includegraphics[width=6cm]{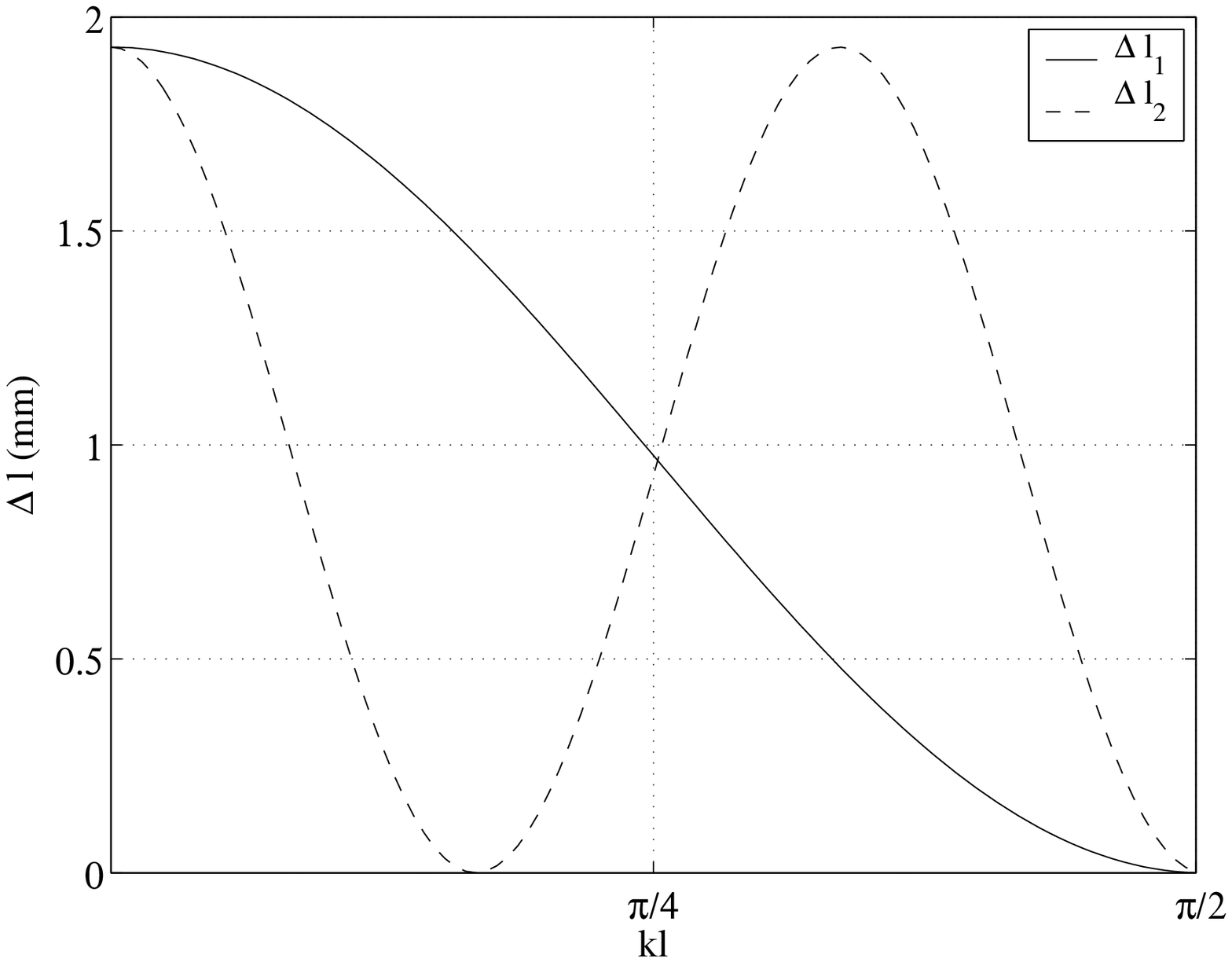} 
\includegraphics[width=6cm]{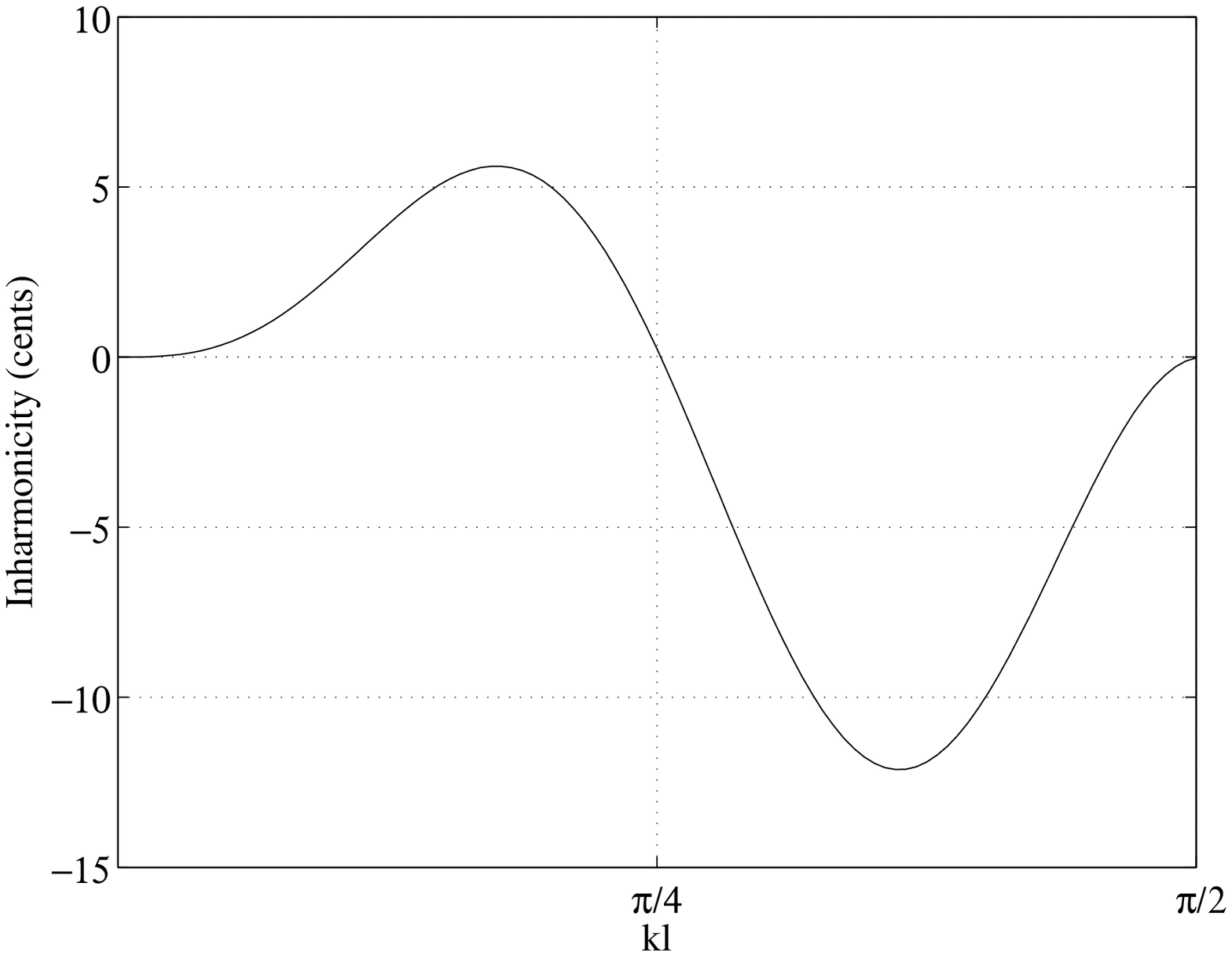}
\caption{\textit{Length corrections (left) and corresponding inter-resonance inharmonicity
(right) for a closed hole whose volume $v=0.3\;cm^{3}$ as a function of $k_1\ell$.}}
\label{fig:inharmo_cav_th}
\end{figure}

\paragraph*{Closed hole (or cavity)}
According to \cite{Dubos:99}, the series and shunt impedances for a closed hole are
\begin{equation}
Z_s=\frac{\rho c}{S_h}[-j \cot k(h+h_m)+jkh_s]\;,
\end{equation}
and 
\begin{equation}
Z_a=j\omega\rho \ell_a^{(c)} /S\;,
\label{eq:Zac}
\end{equation}
respectively, where $\ell_a^{(c)}$ is the series equivalent length for a closed tube given by
\begin{equation}
\ell_a^{(c)}=-r\delta^2[1.78\cotanh(1.84(h+h_m)/r)+0.940+0.540\delta+0.285\delta^2]^{-1}\;.
\end{equation}
In the low frequency approximation, the shunt impedance becomes $$Z_s=\frac{\rho c}{S_h}\frac{-j}{k(h+h_m)}\;,$$
and shows that it behaves as a shunt acoustic compliance $C={v/\rho c^{2}}$ determined by its volume $v=S_h(h+h_m)$. Substituting $Z_a$ and $Z_s$ in equation (\ref{eq:total_tonehole}) and denoting $X=v/S\ell$, the length correction to first order is found to be:
\begin{equation}
\Delta \ell \simeq X\ell\ \cos ^{2}k\ell + \ell_a^{(c)}\sin^2 k\ell\;,  
\label{eq:deltaL_cav1}
\end{equation}
which can be rewritten in
\begin{equation}
\Delta \ell \simeq (X\ell - \ell_a^{(c)})\cos ^{2}k\ell + \ell_a^{(c)}\;.  \label{eq:deltaL_cav}
\end{equation}
This expression shows that the effect of a closed hole inserted on a cylindrical tube depends mainly on its size and position. It also shows that the change in inertance described by $\ell_a^{(c)}$ appears in the determination of the series closed-hole length correction. The pratical magnitude of this term on the total closed-hole effect is to an amount up to $30\%$ for the bottom holes of a clarinet; this cannot be neglected as it could be found in several papers \cite{Keefe:82b,Meynial:88}. This term must be taken into account especially when investigating inharmonicity of a real instrument for which many cavities can act for a given fingering. Since $k\ell \ll 1$, the effect of a cavity is proportional to the ratio of the inserted volume to the volume of air included between the reed tip and the closed hole. As a consequence, the virtual volume of air corresponding to the flow induced by the reed movement introduced by Nederveen, does not produce inharmonicity.  With the use of equation (\ref{eq:deltaL_cav}), inter-resonance inharmonicity becomes:
\begin{eqnarray}
\Delta \ell _{1}-\Delta \ell _{2} &=&-(\frac{v}{S}-\ell_a^{(c)})(\cos^{2}3k_{1}\ell -\cos
^{2}k_{1}\ell )  \notag \\
&=&2(\frac{v}{S}-\ell_a^{(c)})\sin ^{2}2k_{1}\ell \cos 2k_{1}\ell \;,  \notag
\end{eqnarray}
so that:
\begin{equation}
\IH\simeq \frac{4}{\pi }(\frac{v}{S}-\ell_a^{(c)})\,k_{1}\,\sin ^{2}2k_{1}\ell \cos
2k_{1}\ell \;.  \label{eq:inharmo_cav}
\end{equation}
The result depends on two geometrical parameters, the location $\ell$ and the ratio ${v/S}$. Dealing with the case of real clarinet resonator and ignoring the effect of $\ell_a^{(c)}$ can undersestimate the effect of cavities to about 10 cents for the lowest fingerings since each cavity contribution is added. This confirms the necessity to take into account the series closed-hole effect. Besides, as a consequence of the term $\cos 2k_{1}\ell$, either negative or positive inharmonicity is associated with a closed hole (see figure \ref{fig:inharmo_cav_th} for a hole volume equal to $v=0.3\;cm^{3}$). Finally, since the magnitude is proportional to the wavenumber $k_1$, inharmonicity associated with a hole of fixed size and position increases with the fundamental frequency of the played tone. 

\subsubsection{Localized enlargement/contraction}
\begin{figure}[hbtp]
\centering
\includegraphics[width=6cm]{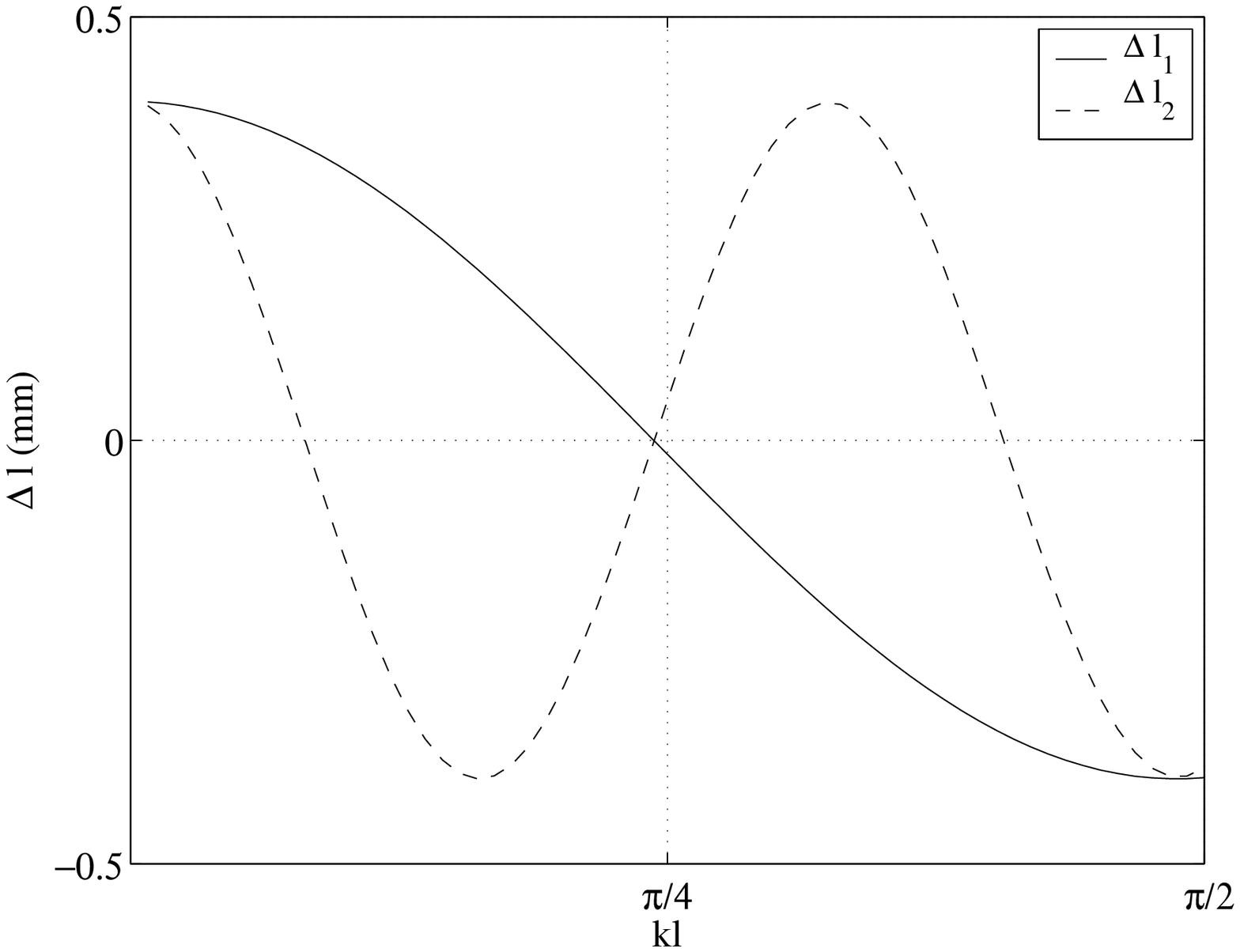} %
\includegraphics[width=6cm]{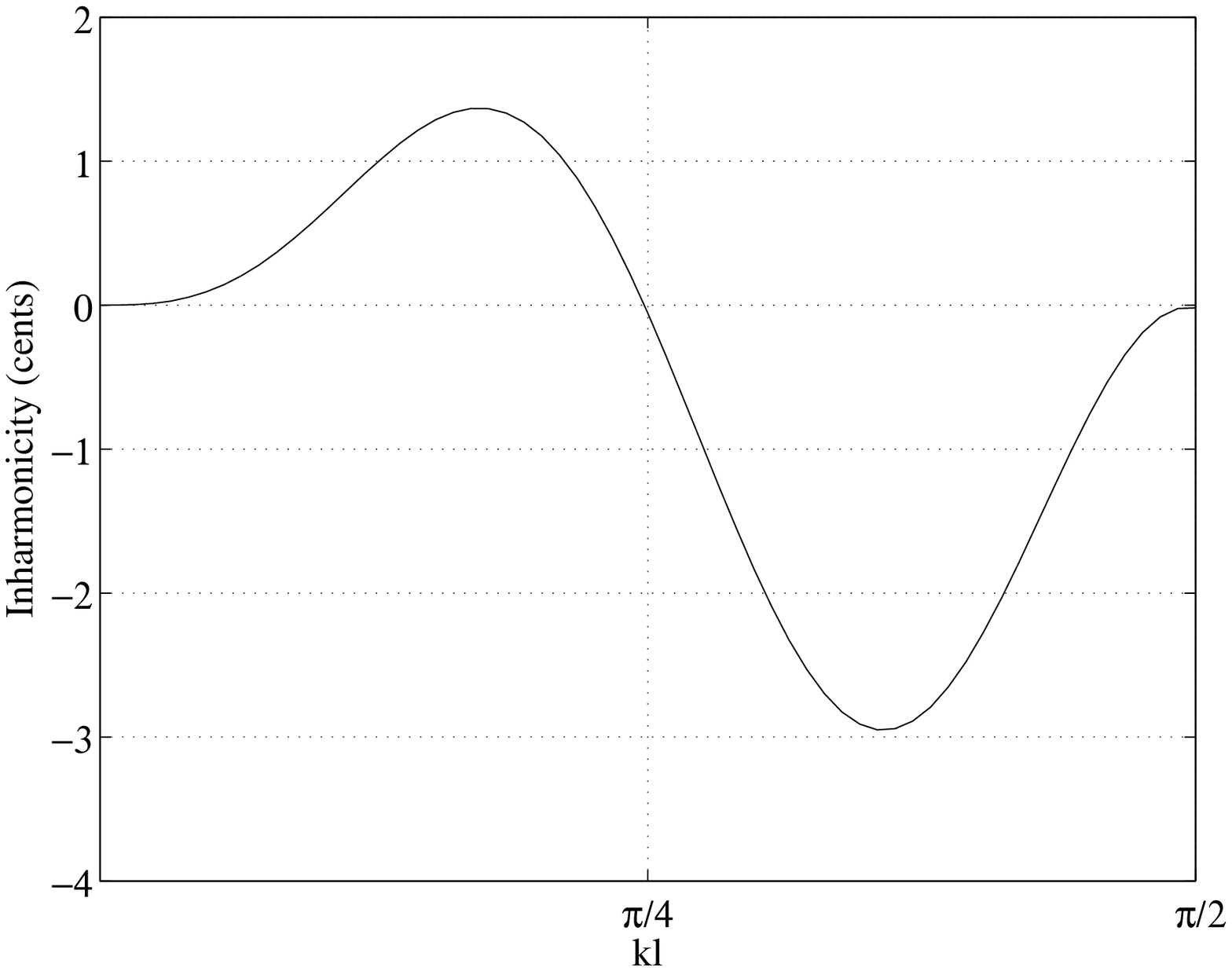}
\caption{\textit{Length corrections (left) and corresponding inter-resonance inharmonicity (right) for a
localized enlargement as a function of $k_1\ell$ ($\ell^{\prime }=10\;mm$, $S^{\prime }=1.04\,S
$).}}
\label{fig:inharmo_disc_loc}
\end{figure}

Consider a localized enlargement (or contraction) of length $\ell ^{\prime }$
located at distance $\ell $ from the effective input in a cylindrical air column and
let $\alpha =S^{\prime }/S$ be the ratio of the cross section area of the
enlargement (or contraction) to the one of the main tube. Assuming $\alpha $ to be close to the unity (only small bore changes are considered), we obtain the change in resonance frequencies expressed through length correction according to (see appendix \ref{anx:un} eq.(\ref{eq:deltaL_localized_exact})):
\begin{equation}
k\Delta \ell =(\alpha-1)\sin k\ell ^{\prime }\cos 2k(\ell +\ell ^{\prime})\;.  \label{eq:deltaL_disc_loc}
\end{equation}
Then, the relationship between first and second resonance frequencies is:
\begin{equation}
\IH\simeq \frac{4}{3\pi }(\alpha -1)(\sin ^{3}2k_1(\ell+\ell')-\sin ^{3}2k_1\ell)\;,  \label{eq:inharmo_disc_loc}
\end{equation}
which states that the two diameter discontinuities  located at points $\ell $
and $\ell +\ell ^{\prime }$ create either positive or negative inharmonicity. This is shown in figure \ref{fig:inharmo_disc_loc} in the case of an enlargement ($\alpha=1.04$) with a $10\;cm$-long inserted tube.

\begin{figure}[hbtp]
\centering
\includegraphics[width=6cm]{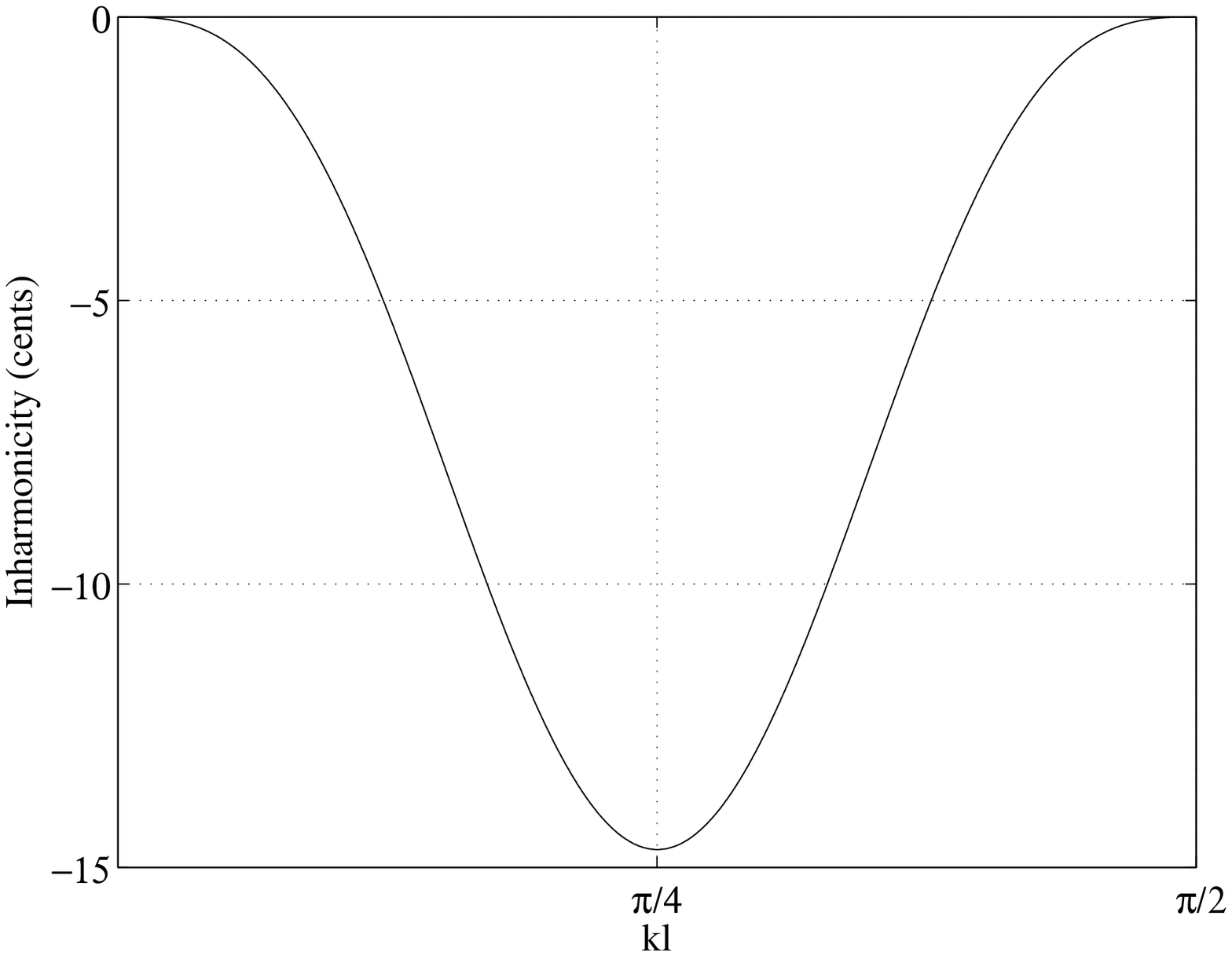} %
\includegraphics[width=6cm]{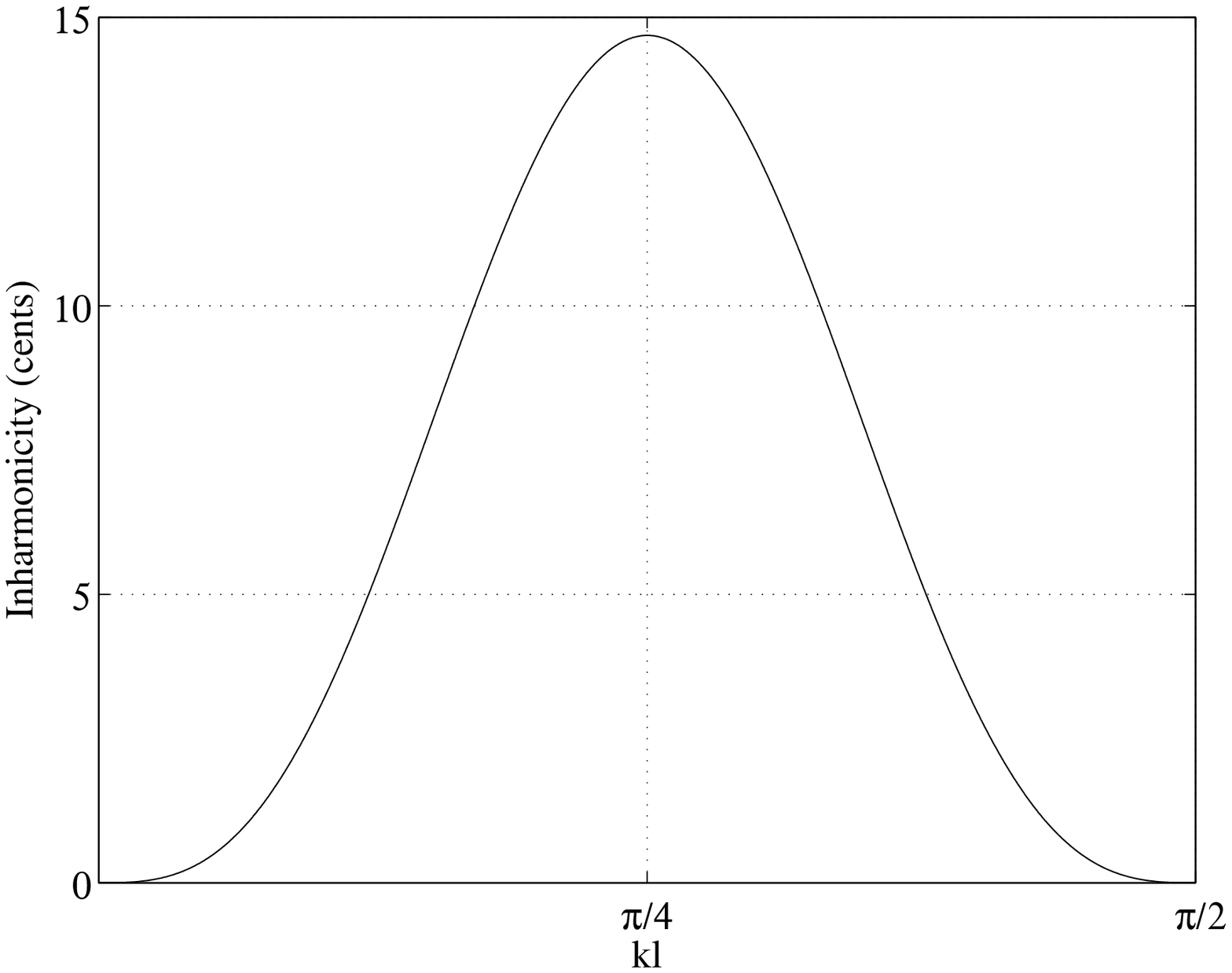}
\caption{\textit{Inter-resonance inharmonicity due to an abrupt change in cross section area as a function of $k_1\ell$\,: contraction $S^{\prime }=0.98\,S$ (left) and enlargement $S^{\prime}=1.02\,S$ (right).}}
\label{fig:inharmo_disc_th}
\end{figure}

\paragraph*{}Substituting $\ell=0$ in equations (\ref{eq:deltaL_disc_loc}) and (\ref{eq:inharmo_disc_loc}), the effects due to a change in cross section in part of a cylindrical tube are calculated. The length correction becomes 
\begin{equation}
k\Delta \ell \simeq {\frac{1}{2}}(\alpha -1)\sin 2k\ell \;,
\label{eq:deltaL_disc}
\end{equation} 
and the inharmonicity can be written
\begin{equation}
\IH\simeq \frac{4}{3\pi }(\alpha -1)\sin ^{3}2k_{1}\ell \;,
\label{eq:inharmo_disc_th}
\end{equation}
where the term $\sin 2k_{1}\ell $ still remains positive since $k_1\ell \in \lbrack 0\;,\pi /2]$. As a consequence, the sign of inharmonicity
associated with a discontinuity depends only on the value of $\alpha $ as shown in figure \ref{fig:inharmo_disc_th} for a bore widening and contraction of 2\%. 

\subsubsection{Change in taper close to the tube input}
\label{subsub:cone} The acoustical behaviour of a change in taper over a
length $\ell _{c}$ can be represented with an equivalent electrical
circuit including two  inductances of opposite sign and the elements of a cylindrical tube of length $\ell _{c}$ \cite{Benade:88bis}. Writing $X_{1}=\ell /x_{1}$  and $X_{2}=(\ell +\ell _{c})/x_{2}$, we obtain the length correction calculated to first order in $X_i$ ($i=1,2$) (see appendix \ref{anx:un} eq.(\ref{eq:deltaL_exact_dv}))
\begin{equation}
k\Delta \ell \simeq X_{2}{\frac{\cos ^{2}k(\ell +\ell _{c})}{k(\ell +\ell
_{c})}}-X_{1}{\frac{\cos ^{2}k\ell }{k\ell }}\;.  \label{eq:deltaL_cone}
\end{equation}
Equation (\ref{eq:deltaL_cone}) states that a single taper change is equivalent to two open side holes with a positive and a negative inertance respectively. It is valid either for a positive or negative taper change, the difference being in the sign of the $x_{i}$ which are positive for a diverging cone and negative for a converging cone. 

\begin{figure}[hbtp]
\centering
\includegraphics[width=6cm]{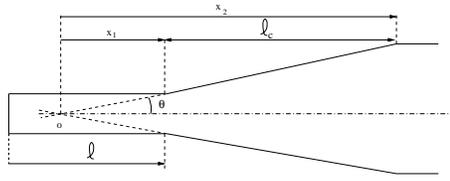} 
\caption{\textit{Geometry and symbol used for the case of a truncated cone.}}
\label{fig:eleccone}
\end{figure}

\begin{figure}[hbtp]
\centering
\includegraphics[width=12cm]{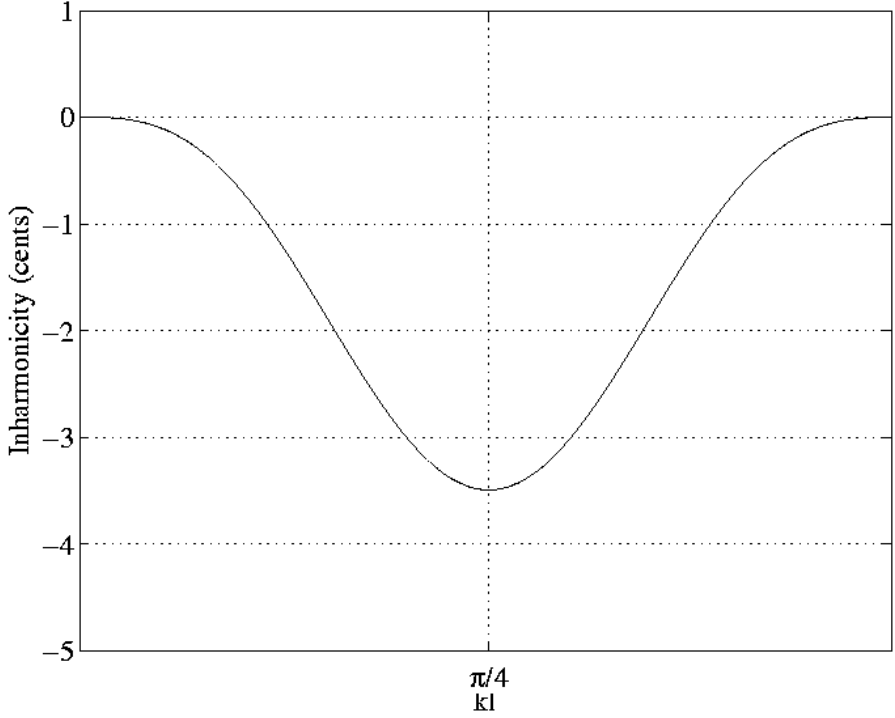} 
\caption{\textit{Inter-resonance inharmonicity for a
positive truncated cone as a function of $k_1\ell$ ($R=7.5\;mm$, $\theta=1.7^\circ$, $\ell_c=5\;mm$).}}
\label{fig:inharmo_cone_bis}
\end{figure}

\paragraph*{}In order to evaluate the inharmonicity generated by a small truncated cone, it
is convenient to reformulate the expression of the length correction in
terms of two control parameters by rewriting equation (\ref{eq:deltaL_cone}). Since approximations $k\ell _{c}\ll 1$ and $\ell _{c}/x_{1}\ll 1$ are still valid, it is possible to write:
\begin{equation*}
\cos ^{2}k(\ell +\ell _{c})= \cos ^{2}k\ell -k\ell _{c}\sin 2k\ell +o(k\ell _{c})\;,
\end{equation*}
and
\begin{equation*}
{\frac{1}{kx_{2}}}={\frac{1}{kx_{1}(1+\ell _{c}/x_{1})}}\simeq \frac{1}{%
kx_{1}}\;.
\end{equation*}
Therefore a simplified expression for the length correction is derived
\begin{equation}
k\Delta \ell \simeq -\frac{\ell _{c}}{x_{1}}\sin 2k\ell \;,
\label{eq:deltaL_cone_modif}
\end{equation}
where $\ell _{c}/x_{1}$ and $\ell $ are the two parameters. Under these
conditions and with the use of equation (\ref{eq:deltaL_cone_modif}),
inharmonicity is given by:
\begin{equation}
\IH\simeq -\frac{8}{3\pi }\frac{\ell _{c}}{x_{1}}\sin ^{3}2k_{1}\ell \;.
\label{eq:IH_troncated_cone}
\end{equation}
An example of diverging cone of length $\ell_c=5\;mm$, large-end radius $R=7.5\;mm$ and half-angle $\theta=1.7^\circ$ is shown in figure \ref{fig:inharmo_cone_bis}. Finally, looking at equation (\ref{eq:deltaL_cone_modif}), the equivalence between a positive truncated cone and an abrupt change in cross section for the case $S^{\prime}<S$ can be noticed. 

\subsection{Other effects: radiation, dispersion and temperature} \label{par:dispersion}
Because of dispersion due to visco-thermal effects, the eigenfrequencies of
the cylindrical air column cannot be exactly harmonically related. From the well known expression of the speed of sound with respect to frequency \cite{Bruneau:98}, it can be shown that dispersion introduces a  positive inter-resonance inharmonicity given by
\begin{equation}
\IH=\frac{\Gamma _{1}-\Gamma _{3}}{1-\Gamma _{1}}\;,  \label{eq:disp}
\end{equation}
where $\Gamma _{n}={\frac{1}{R\sqrt{2k_{n}}}}(\sqrt{\ell _{v}}+(\gamma -1)%
\sqrt{\ell _{h}})$ is the dispersion factor associated to the $n$th eigenfrequencies, $\ell_v$ and $\ell_h$ are the viscous and thermal characteristic lengths and $\gamma$ the ratio of specific heats.  The order of magnitude is given in the next section.\newline
Another effect which affects the relationship between the resonance frequencies is the axial temperature drop. With the relation $\frac{\Delta T}{T}=-\frac{\Delta \rho}{\rho}$, the temperature gradient 
can be seen as a small perturbation in series which modifies the air density. Looking at a location $x$, the infinitesimal length correction is 
\begin{equation}
\delta x = -j Z/Z_c \sin^2 kx \;,
\label{eq:temp}
\end{equation}
where $Z=j\omega\Delta \rho dx/S$. The total length correction is obtained by an integration over the length of the perturbed part of the tube. Assuming arbitrarily a linear temperature profile over the upper third of the length  $\ell_{eff}$ of the instrument  $T(x)=3\;\frac{T_{out}-T_{in}}{\ell_{eff}}x+T_{in}$, where $T_{in}$ is the temperature at the input ($x=0$) and $T_{out}$ is the external temperature, the effect of the thermal gradient is evaluated as follows:
\begin{equation}
\Delta \ell=-3\frac{T_{out}-T_{in}}{T_{out}}\int_0^{x_0} x \sin^2 kx\;dx + \frac{T_{in}-T_{out}}{T_{out}}\int_0^{x_0} \sin^2 kx\;dx \;,
\label{eq:temp_2}
\end{equation}
where $x_0=\ell_{eff}/3$. Expression (\ref{eq:temp_2}) can be calculated analytically since the lower and upper bounds are given by $k_1\ell=\pi/6$ and $k_2\ell=\pi/2$ and approximation for inharmonicity is given by
\begin{equation}
IH=-0.0422\;\frac{T_{out}-T_{in}}{T_{out}}\;.
\end{equation} 
With these assumptions, the thermal gradient appears to produce a constant 5-cent positive inharmonicity across the complete register. Applying a linear thermal gradient over the entire instrument instead of the third of its effective length, would lead to inharmonicity of +10 cents . Finally, with an arbitrary linear temperature profile which is surely not very realistic, it can be said that the thermal gradient have a significant effect on inharmonicity but no more than 5 and 10 cents over the entire scale.\newline
Since radiation of wind instruments depends on frequency, radiation
is also a cause of inharmonicity but its effect is very small compared to the previous effects (less than 1 cent).

\subsection{Theoretical analysis of a clarinet resonator}
\begin{figure}[hbtp] 
\centering
\includegraphics[width=12cm]{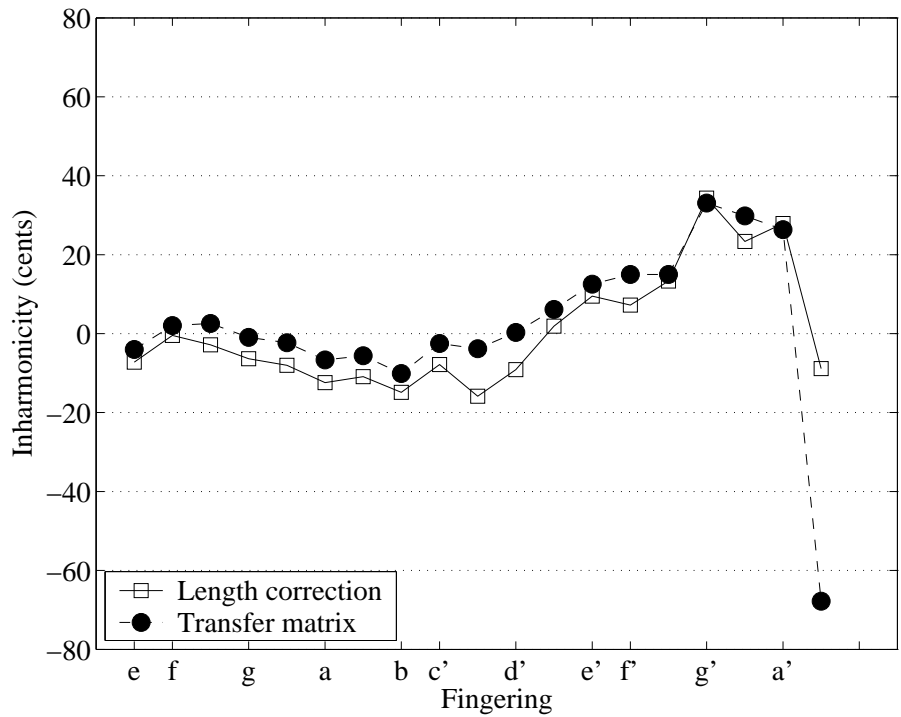}
\caption{\textit{Total inter-register inharmonicity for a Buffet Crampon clarinet. Length correction calculations ($\square$) and transfer matrix calculations ($\bullet$) (the temperature gradient has been ignored).}}
\label{fig:diff_total}
\end{figure}

Our major results concerning inharmonicity are summarized in table \ref{tab:recapitulatif}. 
\begin{table}[hbtp]
\centering
\begin{tabular}{|c|c|c|}
\hline
Basic perturbation & Sign of $\IH$ & Effective parameters\\
\hline
tone hole & $<0$ & $Sh'/S_h$ and $\ell_d$   \\
\hline
closed-hole & $\begin{array}{lcr}
k\ell\lesssim\pi/4 & \Rightarrow & \IH>0 \\
k\ell\gtrsim\pi/4 & \Rightarrow & \IH<0 \\
\end{array}
$ & $v/S$ and $\ell$ \\
\hline
abrupt change in cross section  & $S^\prime>S \Rightarrow \IH>0$ & \raisebox{-.2cm}{$S^\prime/S$ and $\ell$} \\
in the upper part of the instrument  & $S^\prime<S \Rightarrow  \IH<0$ 
 &  \\
\hline
localized enlargement/contraction  &  & \\
in the upper part of the instrument  & \raisebox{+.5cm}{$>0$ or $<0$} & \raisebox{+.5cm}{$S^\prime/S$ and $\ell$}  \\
\hline
diverging truncated cone & & \\
in the upper part of the instrument  & \raisebox{+.5cm}{$<0$} & \raisebox{+.5cm}{$\ell/x_1$ and $\ell$}\\
\hline
converging truncated cone  & & \\
in the upper part of the instrument  & \raisebox{+.5cm}{$>0$} &  \raisebox{+.5cm}{$\ell/x_1$ and $\ell$}\\
\hline
\hline
register hole & $>0$ & $Sh'/S_h$ and $\ell$   \\
\hline
\end{tabular}
\caption{\textit{Sign of inharmonicity associated with basic acoustic perturbations close to the tube input.}}
\label{tab:recapitulatif}
\end{table}
To compare them with a real clarinet, a Buffet Crampon clarinet has been investigated theoretically and experimentally. From the geometrical dimensions of the instrument, given in table \ref{tab:data}, and with the mathematical formulations of inharmonicity between the first two resonance frequencies, both inter-resonance and inter-register inharmonicities were predicted directly for each fingering, simply adding inharmonicities associated to each perturbation. 
\begin{table}[hbtp]
\begin{tabular}{cc}

\begin {tabular}{|c|c|c|c|}
\hline
no &  $\ell$  (mm) & $r$ (mm) & $h$ (mm)\\
\hline
24 & 152.7 & 1.4\phantom{0} & 13.0\\
\hline
23 & 166.9 & 2.35 & \phantom{0}7.1\\
\hline
22 & 193.4 & 2.8\phantom{0} & \phantom{0}7.1\\
\hline
21 & 202.9 & 2.7\phantom{0} & \phantom{0}7.1\\
\hline
20 & 213.2 & 2.55 & \phantom{0}7.2\\
\hline
19 & 230.4 & 2.4\phantom{0}& \phantom{0}6.4\\
\hline
18 & 238.9 & 3.9\phantom{0} & 10.8\\
\hline
17 & 241.3 & 2.35 & \phantom{0}6.8\\
\hline
16 & 252.2 & 2.45 & \phantom{0}8.8\\
\hline
15 & 270.5 & 2.3\phantom{0}& \phantom{0}6.9\\
\hline
14 & 284.3 & 3.3\phantom{0} & \phantom{0}8.8\\
\hline
13 & 287.1 & 2.7\phantom{0} & \phantom{0}6.8\\
\hline
12 & 289.3 & 2.85 & \phantom{0}6.9\\
\hline
11 & 308.2 & 3.95 & \phantom{0}7.4\\
\hline
10 & 318.8 & 2.4\phantom{0} & \phantom{0}7.2\\
\hline
9 & 348.9 & 3.55 & \phantom{0}6.5\\
\hline
8 & 364.8 & 4.05 & \phantom{0}8.9\\
\hline
7 & 369.5 & 3.95 & \phantom{0}6.1\\
\hline
6 & 391.0 & 3.85 & \phantom{0}8.9\\
\hline
5 & 413.1 & 4.85 & \phantom{0}8.9\\
\hline
4 & 445.6 & 5.4\phantom{0}& \phantom{0}6.0\\
\hline
3 & 472.9 & 6.1\phantom{0}& \phantom{0}5.0\\
\hline
2 & 504.9 & 5.6\phantom{0} & \phantom{0}5.2\\
\hline
1 & 543.7 & 5.8\phantom{0} & \phantom{0}4.4\\
\hline
\end{tabular}

 &
\begin {tabular}{|c|c|}
\hline
position ($mm$) & diameter ($mm$)  \\
\hline
0 & 0\\
\hline
89\phantom{0}  &  14.8 \\
\hline
117.5 & 14.6\\
\hline
117.5 & 15.6 \\
\hline
152.5 & 14.5\\
\hline
492.5 & 14.5\\
\hline
526\phantom{0} & 15.1 \\
\hline
565\phantom{0} & 16.9\\
\hline
578.2 & 18.4\\
\hline
598\phantom{0} & 22.2\\
\hline
664\phantom{0} & 60\phantom{0}\\
\hline
\end{tabular}
\end{tabular}
\caption{\textit{Numerical data concerning the Buffet Crampon clarinet investigated. Tones holes are numbered from 1 to 24 in order of decreasing distance from mouthpiece, the $24^{\textit{th}}$ being the register hole. Distances are given from the reed tip. For the input impedance measurements the mouthpiece is replaced by a cylindrical tube of length $77\;mm$. For the comparison with computations, a position correction of about $-12\;mm$ is introduced for the tone hole locations.}}
\label{tab:data}
\end{table}
In order to validate these calculations, computations of the input impedance of the clarinet were carried out using a transmission-line model and transfer matrices. Multiplying sequentially the transfer matrix of each element from the open end to the mouthpiece results in the input response of the instrument. From the obtained impedance curves, inharmonicities between the resonance frequencies were determined for all fingerings. \newline
Calculations with both transfer matrix and length correction approximate formulae coincide satisfactorily for the total inter-register inharmonicity as shown in  figure \ref{fig:diff_total}. Focusing on the influence of a single perturbation, a good agreement is also obtained even if the two methods do not coincide for few tones. For a detailed validation of length correction calculations, readers are referred \cite{Debut:04}. Since the two methods give similar results, our method based on adding corrections to first order seems appropriate.

\subsubsection{Calculation procedure and results}  \label{subsub:calculations}
\begin{figure}[hbtp] 
\centering
\includegraphics[width=12cm]{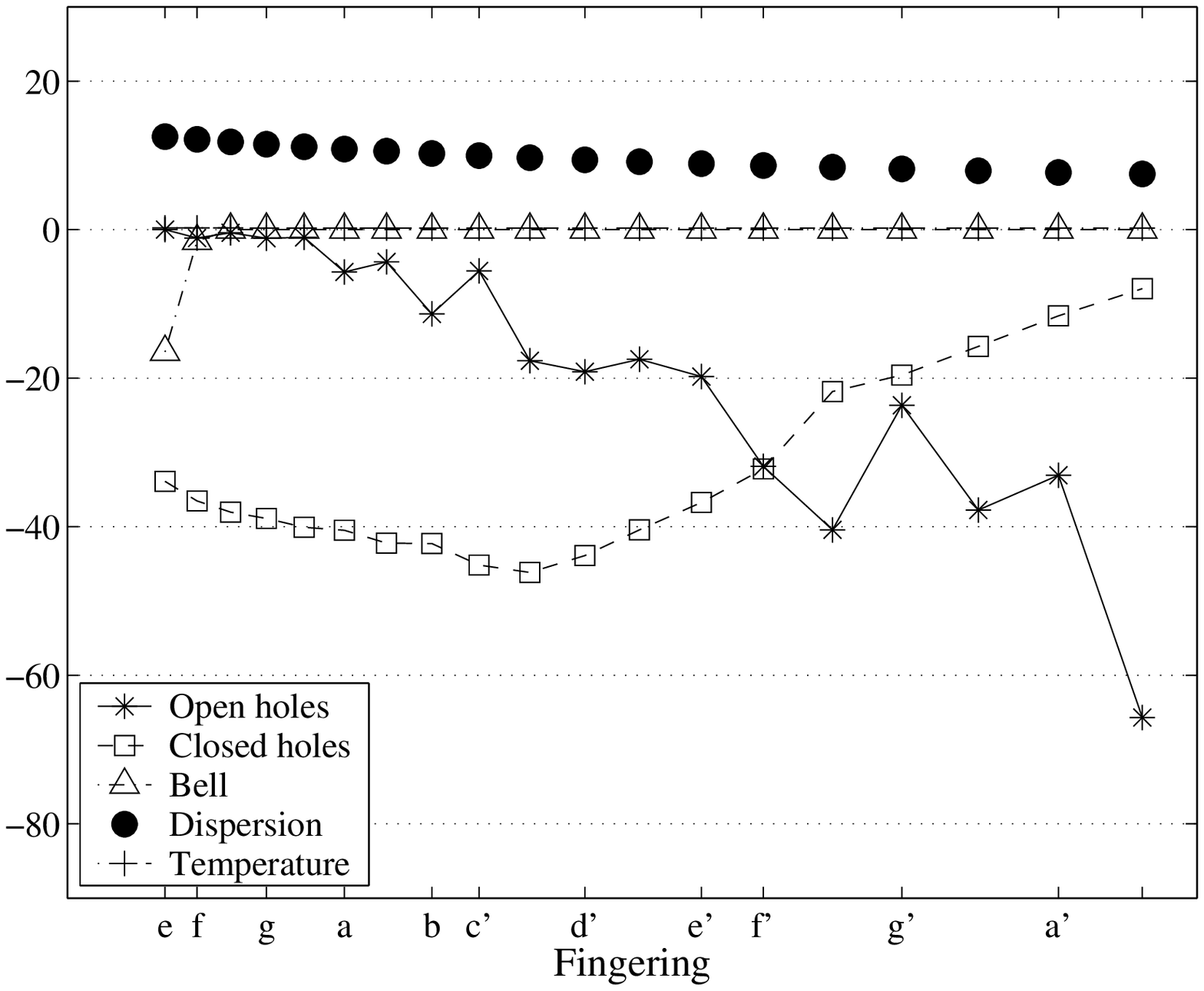}
\caption{\textit{Inter-resonance inharmonicities due to open holes ($\ast$), closed holes ($\square$), dispersion ($\bullet$), temperature ($+$) and the flaring bell ($\triangle$).}}
\label{fig:total_IH}
\end{figure}

\begin{figure}[hbtp]
\centering
\includegraphics[width=12cm]{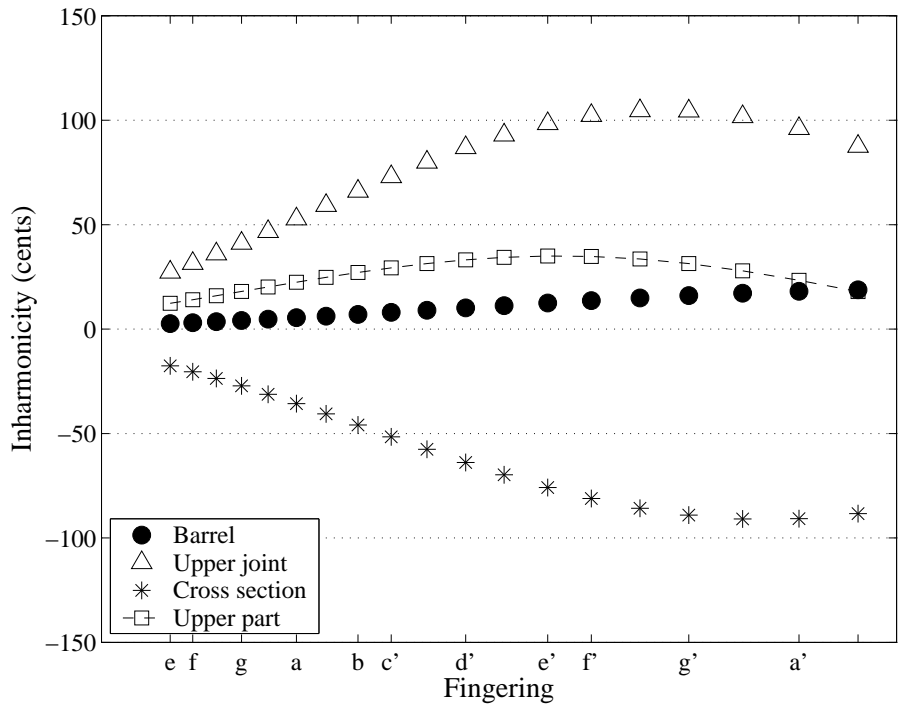}
\caption{\textit{Inter-resonance inharmonicities due to bore perturbations of the upper part: barrel ($\bullet$), change in cross sections ($\ast$), upper joint ($\triangle$) and upper part (sum of the three curves) ($\square$).}}
\label{fig:bore_input}
\end{figure}

The influence on inharmonicity of each bore perturbation found on a clarinet is now discussed and results are plotted in figures \ref{fig:total_IH} and \ref{fig:bore_input}. Results for the register hole have been shown in figure \ref{fig:inharmo_trou_registre}. The principles of the length correction calculations are given for every kind of perturbations. 

\paragraph*{Mouthpiece}The acoustical top of the instrument (i.e the effective input) for measurements is defined assuming a cylinder of length $\ell_G=77\;mm$. Even if this can be seen as the sum of an equivalent embouchure of volume $v=11\;cm^3$ plus a length correction due to reed effects equal to $10\;mm$ as measured in reference \cite{Dalmont:95} by comparing the resonance and playing frequencies, it must be noticed that this equivalent length is imposed by the experimental setup and ignores the variation of reed effects with frequency. This characterisation of the mouthpiece does not take into account the geometry of the input volume which influences the functioning. Actually, the theoretical analysis is made in two steps:
\begin{description}
\item i) the analysis of the input impedance with the equivalent embouchure (present section \ref{subsub:calculations} );
\item ii) after a comparison of the results of the present section with experiments, a discussion of the effect of the conicity of the mouthpiece is given in the next section (see \ref{sub:comparison}).
\end{description} 
\paragraph*{Barrel and top part of the upper joint}
Both the barrel and the top part of the upper joint are tapered towards the bottom with a step discontinuity  between them. Inharmonicity due to change in cross-section is evaluated with equation (\ref{eq:inharmo_disc_th}) and since the assumption $X\ll1$ is not true for both changes in taper, inharmonicities are derived using exact formulae for length corrections (see appendix \ref{anx:un} eq.(\ref{eq:deltaL_exact_dv})) and equation (\ref{eq:inharmonicite}).\newline 
It appears that the barrel entails positive inharmonicity and affects much more the highest part of the register than the lower part. Besides, even if the change in cross section tends to balance the effect of the inverse conical upper joint (see figure \ref{fig:bore_input}), the total inharmonicity associated with bore perturbations at the input is positive and appears to be essential to correct the damage caused by open holes, cavities, dispersion and the flaring horn as it will be shown below. \newline
An important point to notice is that since length corrections due to taper and discontinuity in cross section are frequency dependent, a correction is needed for the location of closed tone holes, depending on the considered fingering.

\paragraph*{Open holes}As apparent with equation (\ref{eq:IH_trou_ouvert}), inharmonicity associated with open holes depends strongly on the length $\ell_{d}$ of the tube-part below the first open hole. To allow  realistic calculations, a good estimation of this parameter is nedeed. Instead of approximating $\ell_d$ as the distance between the first two open tone holes, the following method is used. Starting from fingering $f$ with hole 1 open, calculation of $\ell_{d}$ is self-evident: an equivalent length $\Delta \ell$ of the two removed admittances ($Y$ and $Y_{down}$) can be deduced with equation (\ref{eq:deltaL_troudenote}). This gives the substitution tube of length $\ell+\Delta \ell$ to produce tone $f$. Dealing with the next tone $f\#$, $\ell_{d}$ is now approximated by the distance between the first two open holes increased by the tube-piece of length $\Delta \ell$ calculated previously. This allows to take into account, up to a certain extent, the effect of the tone hole lattice (only two or three open holes may play a role in the determination of the playing frequencies \cite{Benade:60}). Then, inharmonicity is deduced for this fingering, a new $\Delta \ell$ is calculated and so on. In practice, the method is very convenient but fails for few situations such as cross-fingering, because of the proximity effects between two open holes. \newline
As it is expected, examination of inharmonicity due to open side holes shows the tendency to produce negative inharmonicity across the entire scale (see figure \ref{fig:total_IH}). Besides, it appears that the highest tones of the register are more affected than the bottom tones because of the strong dependence of inharmonicity with the wavenumber $k_1$ (see equation (\ref{eq:IH_trou_ouvert})). 

\paragraph*{Closed holes}
Inharmonicity due to a closed side-hole located above the first open hole is calculated with equation (\ref{eq:inharmo_cav}) assuming that every hole is independent of the others. Adding each closed hole contribution, the total closed-side hole inharmonicity is determined.\newline
Figure \ref{fig:total_IH} shows the important role of the closed holes. Inharmonicity is negative and the shape of this curve can be interpreted by studying the combined action of cavities located at both extremities of the tube of length $\ell_{eff}$. It must be reminded that inharmonicity due to a closed hole is mainly proportional to the hole volume, the hole volumes being larger for cavities near the open-end than those located at the input of the instrument. On one hand, for tones from $e$ to $c^{\prime}\#$, cavities located near the open-end have a larger (negative) effect than those located near the mouthpiece which produce small positive inharmonicity. As a consequence, harmonicity is more and more altered. On the other hand, for tones from $c^{\prime}\#$ to $a^{\prime}\#$, inharmonicity still remains negative but decreases when the chromatic scale is played which is mainly due to the decreasing of the number of closed holes.

\paragraph*{Flaring horn}The expanding part at the open-end of the clarinet has been modeled as a catenoidal horn with a horn constant $h=0.085$. Using the procedure given in \cite{Nederveen:69} and restricting our investigation to the two lowest tones only, a length correction referring to the geometrical length has been assigned to the flaring bell according to
\begin{equation}
\Delta \ell_{h}=-\ell_{h}+1/k\,\arctan (kh'\;\tan(L_{h}/h')),
\label{eq:deltaL_horn}
\end{equation}
where $L_{h}$ is the effective length including the classical end-correction $0.6R_f$ ($R_f$ being the end-radius of the flare) and $h'=h/\sqrt{1-k^2h^2}$. Besides, a position correction has been given to the two holes located in the expanding part of the bell as Nederveen did. As expected, the flaring horn produces negative inharmonicity for the lowest tone to about -15 cents and has small influence on fingering $f$.

\paragraph*{Dispersion and temperature}
Calculations of the effects of dispersion is obtained with equations (\ref{eq:disp}) and results are plotted in figure \ref{fig:total_IH}. Dispersion effect is  continuous over the entire scale of an amount to about +10 cents. Concerning the effect of the thermal gradient, it has been discussed in section \ref{par:dispersion}.

\subsection{Comparison between theory and measurements}\label{sub:comparison}
\begin{figure}[hbtp] 
\centering
\includegraphics[width=12cm]{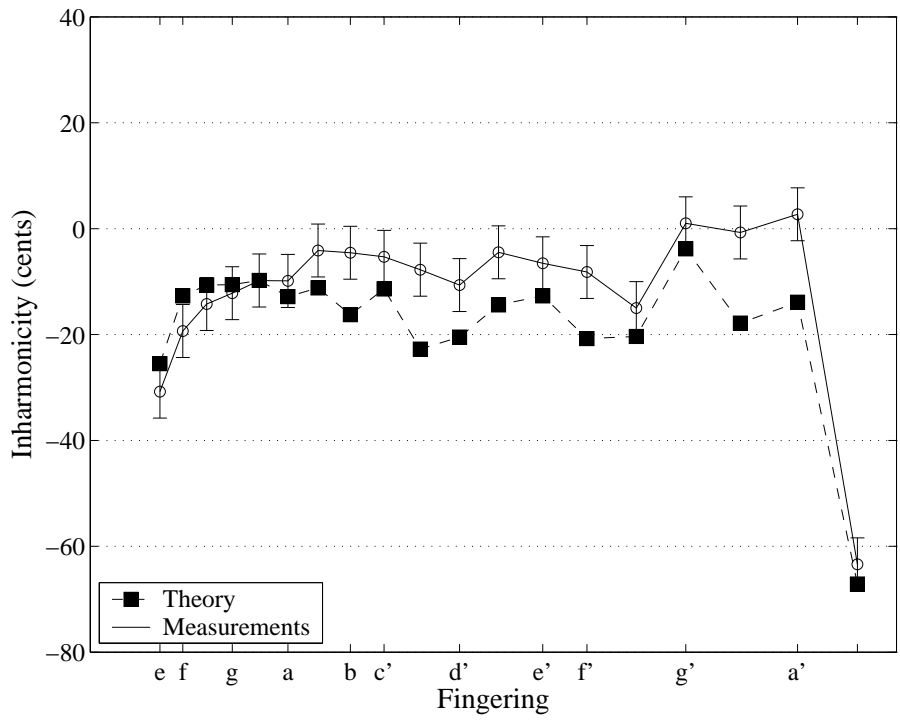}
\caption{\textit{Inter-resonance inharmonicity. Comparison between measurements (error bars) and length correction calculations ($\blacksquare$) (uniform temperature).}}
\label{fig:mes_ss_reg} 
\end{figure}

\begin{figure}[htbp] 
\centering
\includegraphics[width=12cm]{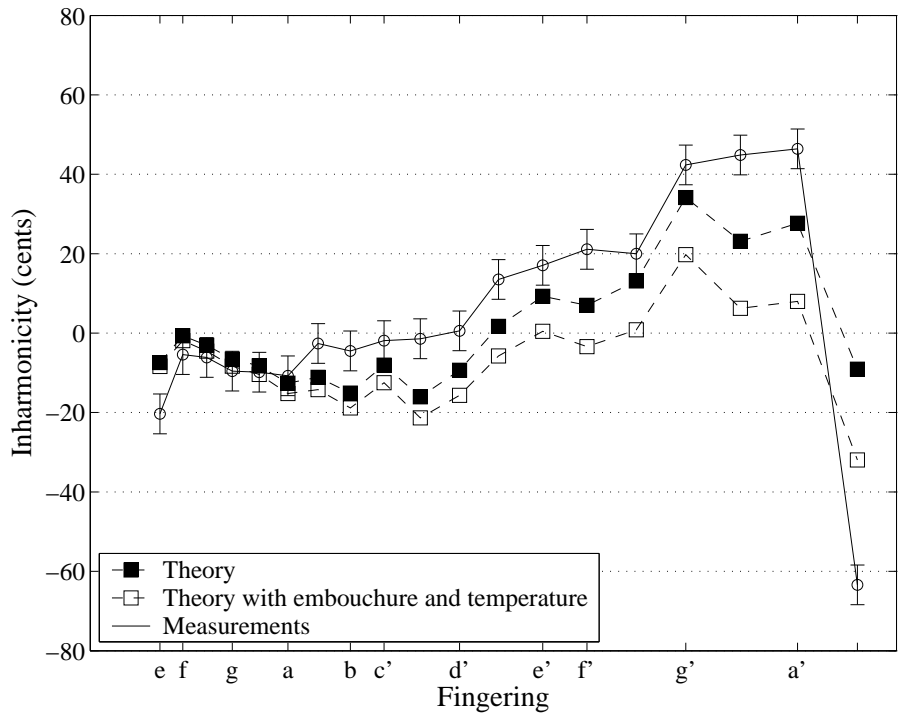}
\caption{\textit{Inter-register inharmonicity. Comparison between measurements (error bars) and length correction calculations ($\blacksquare$). Adding the effects of the thermal gradient and the conicity of the embouchure to ($\blacksquare$) gives ($\square$).}}
\label{fig:mes_reg} 
\end{figure}

\begin{figure}[htbp] 
\centering
\includegraphics[width=12cm]{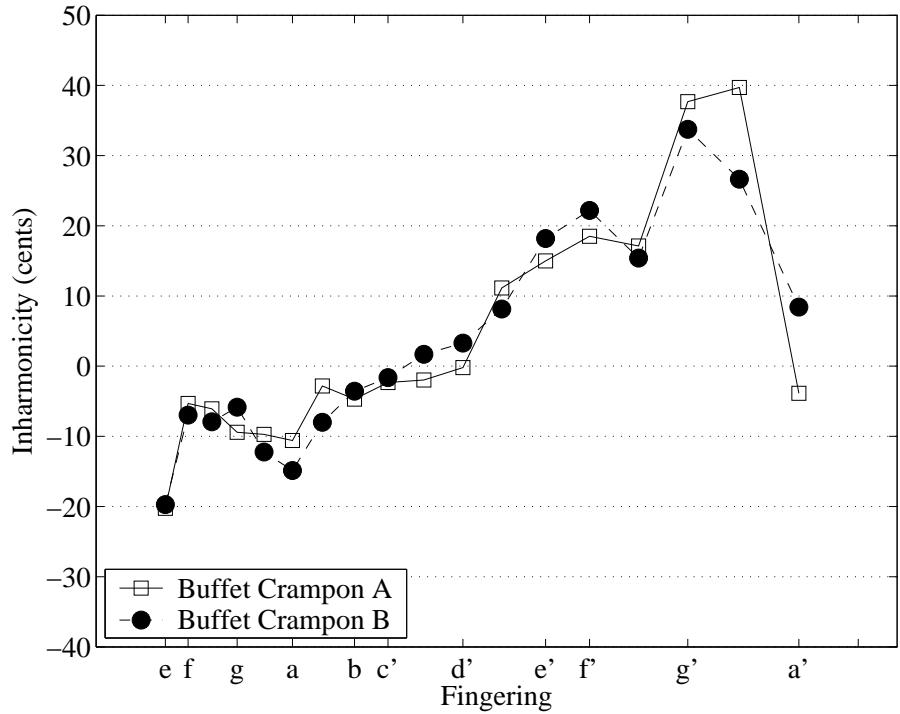}
\caption{\textit{Measured inter-register inharmonicities for two Buffet Crampon clarinet models.}}
\label{fig:two_clarinets} 
\end{figure}

To correlate theoretical results and measurements, an input impedance measurement device is used (see reference \cite{Dalmont:92}). It gives the acoustic linear response of the passive instrument (i.e without reed effects) to an harmonic excitation. Measurements are made in a well-insulated room and temperature is $18.5^{o}C$ \footnote{Two other changes in cross section occur in the experimental setup at the input of the instrument but their effect on the resonance frequencies is substracted algebraically.}. The resonance frequencies are obtained by interpolating the imaginary part of the admittance in the vicinity of each resonance, the oscillating frequency satisfying ${\mathcal{I}}m(Y)=0$. The absolute uncertainty on the measurements are determined to be $\pm5\,$ cents.\newline
Figures \ref{fig:mes_ss_reg} and \ref{fig:mes_reg} show that experiment and theory are in a good agreement, except for few tones for which discrepancies can be explained by an inconvenient determination of the length $\ell_d$ when investigating the open hole effect. 
Figure \ref{fig:mes_reg} also shows the inter-register inharmonicity with conditions close to the normal use of a clarinet, i.e when both the thermal gradient and the conicity of the embouchure at the end of the mouthpiece are taken into account. With the use of the geometry of a classical embouchure (see reference \cite{Vergez:03}), the effect of the conical part at the end of the embouchure has been evaluated with the exact formulae of length corrections given in appendix \ref{anx:un} (see eq.(\ref{eq:deltaL_exact_dv})). It appears that the embouchure conicity narrows the inharmonicity of the highest tones of the register. \newline
Otherwise, two clarinets were investigated experimentally. It appears that both clarinets have the same general tendencies in inharmonicity as figure \ref{fig:two_clarinets} shows, even if small differences can be noticed for few tones. This fact has previously been reported in \cite{Dalmont:95} and is linked to differences in bore profile of the upper part of the instrument. In the same reference, figure 11 shows results for two different clarinets,
very similar to the results of figure \ref{fig:two_clarinets} for the inter-resonance inharmonicity:
nevertheless a tendency for the results of the present study is to be 5 cents lower than the results for the clarinets measured in \cite{Dalmont:95}. Moreover, adding the results for the inter-resonance inharmonicity and for the inharmonicity due to register hole leads to very similar inter-register inharmonicity results for the two studies.

\subsection{Discussion and conclusions}
\begin{figure}[htbp] 
\centering
\includegraphics[width=12cm]{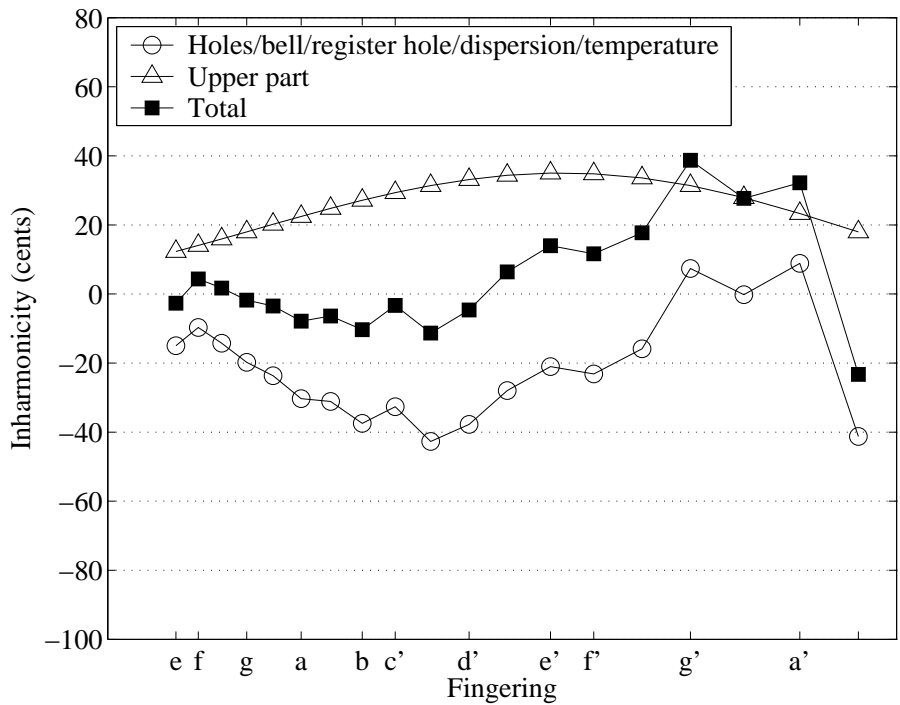}
\caption{\textit{Inharmonicities due to perturbations found in the upper part ($\triangle$), due to hole/bell/register hole/temperature/dispersion ($\circ$), and resulting total inter-register inharmonicity (sum of the two curves) ($\blacksquare$).}}
\label{fig:total_reg_IH}
\end{figure}

\paragraph*{i)}All our calculations deals with resonance frequencies and no excitation mechanism has been introduced. As shown by figures \ref{fig:total_IH} and \ref{fig:total_reg_IH}, our analysis leads to the following statements concerning the origin of the general tendency of clarinets to have tuning deficiencies for the lowest twelfths:
\begin{description}
\item - the used linear theory does not reveal a tuning problem for the lowest tones,
\item - certain dimensions of the tone holes have been chosen by makers. When their effects are considered as imposed, the remaining parameters to achieve a correct tuning are mainly in the bore profile and the location of the register hole,
\item - a good compromise between the bell profile and the register hole location and size, is needed to control the tuning of the lowest twelfth (tone $e$), 
\item - the barrel and bore profile at the top end of the upper joint are of great importance in order to bring the lowest and middle twelfths in tune. This effect tends to balance the effect of the closed holes over the entire scale.\end{description}
Finally, if positive inharmonicity of the resonator is observed for the lowest twelfths, the origin of the problem may lie in combination of the combined action of register hole opening, cavities, bell and inappropriate bore perturbations in the upper part. Nederveen \cite{Nederveen:69} has raised the question of the necessity of bore perturbations in musical instruments. In the case of the clarinet, calculations reveal the need to resort to bore irregularities close to the input to compensate for the negative inharmonicity caused by open holes, cavities, flaring bell and dispersion. Reaming the upper bore profile appears to be a difficult task and must be done carrefully, since this part seems to control the accuracy of the twelfths.
\paragraph*{ii)}As discussed in section 2, predicting playing frequencies from resonance curves is not an easy task. Reference \cite{Dalmont:95} shows that the
measured length correction due to the reed effects (volume velocity of the
reed and reed damping) is rather constant over the entire first register and
the first half of the second one, then increases strongly. This allows compensation for the large increase of inter-register inharmonicity shown e.g. in figure \ref{fig:two_clarinets}. Nevertheless this kind of measurement, done using an artificial mouth, is rather difficult. It is also difficult to compare directly to the results obtained at different levels by an instrumentalist. An example of
this difficulty is given in \cite{Bak:87}. From our results for the two clarinets of figure \ref{fig:two_clarinets}, it is not possible to
explain why the players are obtaining too large twelfths for the first tones
($e$ to $g$). It is a way for future investigation: maybe the analysis could
be made using also an artificial mouth, but the fact is that the players
often complain about too large twelfths, and this fact seems to be quite common:
as an example, recently Buffet Crampon has designed a new clarinet, called
\textit{Tosca}, correcting the tone $f$. For these low tones, what is the optimum
inharmonicity of the resonator for having well tuned twelfth? Is it zero? The answer is not
obvious, depending also on the excitation level, and certainly on the
''tunability range'' \cite{Leipp:76,Deplus:private}, due to several factors, including the
modification of both the embouchure (reed opening and reed damping) and the vocal tract. These musician-control parameters have not been extensively investigated yet. Moreover the
''tunability range'' increases with the pitch of the played tone: it is larger
for the higher register than for the lower one, particularly due to the low
level of the second impedance peak (for the second register, the second peak
corresponds to the sixth peak of the first register).
\paragraph*{iii)}In the next sections, we continue our study in the same direction but in a different perspective. First, the register hole is searched to allow a register jump for the 19-tone complete register. This is an alternative to what makers do since clarinettists use different fingerings to play tones from $f^{\prime}\#$ to $a^{\prime}\#$ and their associated twelfths. Second, instead of working with a real instrument, we consider a perfect cylindrical tube (without tone holes and flare) with only one register hole. The next sections are devoted to answer to the following general question:  is it possible to find a register hole location on a cylindrical tube, combined with a simple perturbation in the upper part, which allows for the first two complete registers to be perfectly well tuned? In other words, we examine if it is possible to design a well-tuned instrument which is provided with additionnal bore perturbations in the upper part which exactly compensate for inharmonicities due to tone holes and other effects. 
\section{Optimization of the location of the register hole and tuning corrections : statement of the problem}
\markboth{On the tuning defaults related to the register hole of a clarinet}{On the tuning defaults related to the register hole of a clarinet}
\subsection{Assumptions}
Our goal now is to achieve harmonicity of the first two resonance
frequencies $f_{1}$and $f_{2}$ , for the effective lengths corresponding to
each hole, i.e. between two extreme values of $\ell _{eff}$, written $\ell
_{\min }$ and $\ell _{\max }$. The precise intermediate values of the length
is without importance for the present objective. As a consequence, the
absolute effect of small discontinuities, e.g. cavities or tapers, is not
important as well, only their relative effect between these two registers
being important. It is therefore convenient to consider a continuous
variation of the length between the two extreme values. After the
optimization of the intervals between the two registers between these two
values, it will be possible to find the precise location of the tone holes
achieving the desired scale.
\paragraph*{}How can harmonicity between the frequency $f_{1}$ of the tone of the
first register and the frequency $f_{2}$ of the tone when the
register hole is open, be achieved for all lengths between $\ell _{\min }$ and $\ell
_{\max }$? First we will find an optimal location for the register hole,
located upstream of $\ell _{\min }$, secondly we will study if a correction
system can compensate the residual defaults of this register hole. The
dimensions of the hole are considered to be optimized by the practice of
makers, and therefore are regarded as imposed. It is actually a difficult
question, related to nonlinear effects as well as water effects, the
important fact being that the linear behaviour, at low level, is well known \cite{Berg:01}.
We do not take into account the use of the register hole as a B flat tone hole. The correction systems are sought in order to be without manipulation by the
instrumentalist, and therefore to act on both the first and the second
registers, contrary to the register hole itself.
\subsection{Formulation of the optimization problem}
Optimization techniques are used to find a set of design parameters $x_\star=\{x_\star^{(1)},\dots,x_\star^{(n)}\}$ that can in some way be defined as optimal. These parameters are obtained by minimizing (or maximizing) a criterion function $\ff$ which may be subject to constraints and/or parameter boundaries. Thus, the design parameters are subjected to the following requirements:
\begin{description}
\item - small changes location must be less than the distance between the reed tip and the first tone hole;
\item - dimensions of the acoustical systems must be reasonable for the realization;
\item - geometrical dimensions are positive.
\end{description}
Our optimization problem is  formulated as follows:
\begin{equation}
\left\{
\begin{array}{l}
\d\min_{x\in\RR^n} \quad \ff(x) \quad \hbox{where } \ff \; \hbox{is the criterion function}  \\ 
l_{b_i}\leq x_i \leq u_{b_i} \quad \hbox{i=1,\dots, n}
\end{array}
\right.\label{eq:pb}
\end{equation}
where $x$ is the vector of design parameters ($x\in\RR^n$) and $l_{b_i}$ and $u_{b_i}$ are the lower and upper parameter boundaries respectively, for the design parameter $x_i$.  
\subsection{Criterion functions}
\subsubsection{Location of the register hole}\label{par:opt_position}
As previously shown, the presence of the register hole means that it is probably impossible to built a clarinet in a way that the two registers are played accurately in tune with the same fingerings. Thus, the objective is to find the register hole location that entails the smallest frequency shift of the second resonance frequency for all fingerings. The first optimization is performed in order to play a 19-tone compass i.e a complete register. Then, a second optimization is performed restricting attention to fingerings from \textit{e} to \textit{f'} (above this fingering, clarinettists do not use the same fingering to play the fundamental and the associated twelfth). The distance $\ell$ of the hole from the  effective input is the only design parameter. The optimization problem deals with equation (\ref{eq:inharmo_reg}) which predicts theoretically the frequency deviation associated with the opening of the register hole. It can be formulated with one of the two following criterions: 
\begin{itemize}
\item criterion 1: to minimize the maximum of the frequency deviation;
\item criterion 2: to minimize the mean of the square of the derivative with respect to $k$ of equation (\ref{eq:inharmo_reg}).
\end{itemize} 
The first criterion consists in limiting the most important tuning default of the instrument. This criterion, which is very simple and intuitive for anyone who is interested in instrument design, can be written:
\begin{equation}
\label{eq:cout_reg_max}\hbox{\textbf{criterion 1}:}\quad \ff_1=\sup\left(\frac{2}{3\pi}\d\frac{s}{Sh'}\d\frac{cos^2 k_2\ell}{k_2}\right)\;.\end{equation}An interesting point about criterion 1 is that the solution can be approximated analytically as it is shown in section \ref{par:critere1resu}. On the contrary, once the maximum of the deviation is achieved for a fixed register hole location, the frequency deviations associated with other fingerings are not taken into account for the evaluation of the function.\newline
Concerning criterion 2, the criterion function to minimize is formulated as
\begin{equation}
\label{eq:cout_reg}\hbox{\textbf{criterion 2}:} \quad\ff_2=\d\frac{1}{2}\int_{k_{min}}^{k_{max}}\left[\d\frac{\partial}{\partial k_2}\left(\frac{2}{3\pi}
\frac{s}{Sh'}\frac{cos^2 k_2\ell}{k_2}\;\right)\right]^2 dk_2\;,\end{equation} where $\partial/\partial k_2$ is the derivative with respect to the wavenumber $k_2$, $k_{min}$ and $k_{max}$ being related to the playing frequency via the effective length by $k_{min}=3\pi/2\ell_{max}$ and $k_{max}=3\pi/2\ell_{min}$ respectively. Contrary to criterion 1, this criterion function takes into account the deviation associated with all fingerings for its evaluation. The global minimum of this criterion function is achieved when the integrand is zero, i.e when the frequency deviation is constant for all fingerings. Nevertheless, noting that equation (\ref{eq:inharmo_reg}) is necessarily equal to zero for a particular frequency, $k_2\ell=3\pi/2$, it appears that the register hole location given by this criterion function is the one that minimizes inharmonicity variations around zero and may lead to a more homogeneous register jump across the entire register.     
\subsubsection{Correction of the register hole effect}\label{par:fct}
The second aim of this optimization is to give suggestions on how to compensate for the register hole effect.  We are looking for geometrical dimensions of acoustical systems whose effects alter the resonance frequencies in order to restore the original $f_2/f_1$-mode frequency ratio. Similarly to the study of the location of the register hole and denoting $\IH_{reg}$ and $\IH_{pert}$ the inharmonicities associated with the register hole and the perturbation respectively, a criterion function can be written as 
\begin{equation}
\label{eq:cout_opt_max}
\hbox{\textbf{criterion 1}:\quad} \ff_1=\sup\left(\vert\IH_{reg}+\IH_{pert}\vert\right)\;,
\end{equation}
which deals with the maximum of the total inharmonicity. Moreover, a second criterion function has also been investigated by minimizing the function defined as
\begin{equation}
\hbox{\textbf{criterion 3}:\quad}\ff_3=\frac{1}{2}\int_{k_{min}}^{k_{max}}\left[\IH_{reg}+\IH_{pert}\right]^2dk \;.\label{eq:critere}\end{equation}
for which the minimum is reached when inharmonicity associated with the inserted perturbation is of the same magnitude as the register hole deviation and in the opposite direction.
\section{Results and discussions}\label{sec:resultats}
For the optimization, the radius of the main tube is taken as $R=7.5\;mm$. The height of the register hole is $12.5\;mm$ and its radius $1.55\;mm$. The upper bound for the acoustical system location, is set to $154\;mm$ which is the first tone hole location according to table \ref{tab:data}. The other parameter boundaries for all perturbations have been chosen in order to make the realization possible.   

\subsection{Register hole location}
\subsubsection{Criterion 1: maximum of the frequency deviation}\label{par:critere1resu}
\begin{figure}[h!tpb]
\centering
\includegraphics[width=6cm]{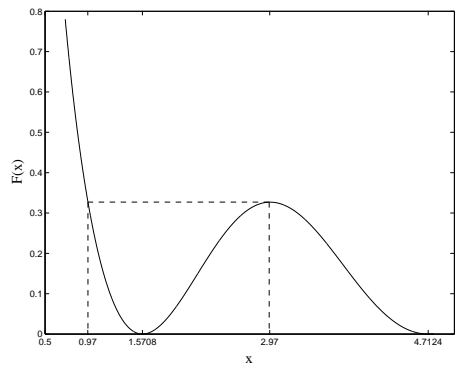}
\caption{\textit{Perturbation function $F(x)=\cos^2x/x$. The same local maximum is reached for $x_0=2.97$ and $x_\star=0.97$.}}
\label{fig:perturbation}
\end{figure}

\begin{figure}[h!tpb]
\centering
\includegraphics[width=6cm]{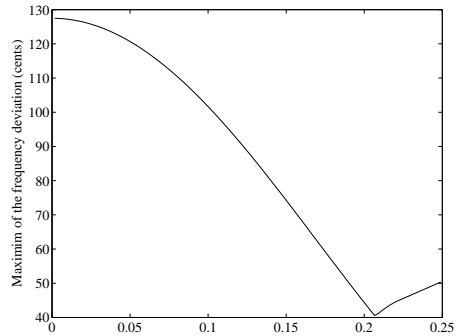}
\caption{\textit{Maximum of the frequency deviation as a function of $\ell/L$. Above $\ell/L=0.2041$, the maximum value of the frequency shift increases linearly with the register hole location.}}
\label{fig:criterion}
\end{figure}

As mentioned earlier, an optimum of criterion 1 defined by equation (\ref{eq:cout_reg_max}) can be derived analytically. In order to achieve the solution, we rewrite the frequency shift as follows
\begin{equation}
\label{eq:IH_trou_bis}
\IH=\d\frac{2}{3\pi}\d\frac{s}{Sh}\;\ell\;\d\frac{cos^2 k_2\ell}{k_2\ell}\;,
\end{equation} 
and note that the magnitude of the deviation is proportional to $\ell$ and varies as the function $F(x)=\cos^2x/x$ where $x=k_2\ell$. We will first proof that the maximum of $F(x)$  does not depend on $\ell$ for a certain interval of values of $k_2\ell$ and therefore the optimum value of $\IH$ is obtained for the minimum of $\ell$ on this interval.\paragraph*{}The effective length of the tube varies between $\ell_{max}=L=585\;mm$, the total length of the instrument including reed and radiation length corrections, and $\ell_{min}$. If the compass of a register is one twelfth minus one semi-tone, the two extreme values of the effective length are related by $\ell_{min}=\ell_{max}/2^{18/12}\,$ which for simplicity we assume first to be $\ell_{min}=\ell_{max}/3$. The location of the register hole, defined by $\ell$, is in the upper part of the instrument and satisfies $\ell<\ell_{min}$.\newline 
Then, for a given effective length, $\ell_{eff}$ lying between $\ell_{min}$ and $\ell_{max}$, the wavenumber is defined as $k_2=3\pi/2\ell_{eff}$, and therefore the argument of $F(x)$ varies follows : 
\begin{equation}
\frac{3\pi\ell}{                                                                                                                                                                                                                                                                                                                                                                                                                                                                                                                                                                                                                                                                                                                                                                                                                                                                                                                                                                                                                                                                                                                                                                                                                                                                                                                                                                                                                                                                                                                                                                                                                                                                                                                                                                                                                                                                                                                                                                                            2\ell_{max}}<k_2\ell<\frac{3\pi\ell}{2\ell_{min}}\;,
\end{equation}
thus
\begin{equation}
\frac{3\pi}{2}\frac{\ell}{L}<k_2\ell <\frac{9\pi}{2}\frac{\ell}{L}\;.
\label{eq:intervalle_k}
\end{equation}
Because the ratio $\ell/L$ is less than 1/3, the upper bound for $x=k_2\ell$ is therefore $3\pi/2$ for which $F(x)=0$. Figure \ref{fig:perturbation} shows the variation of $F(x)$. From $\ell=L/3$, for the interval defined by inequalities (\ref{eq:intervalle_k}), $F(x)$ varies between 0 and 0, with a maximum value equal to $0.327$, for $x_0=2.975$. When $\ell$ decreases from $L/3$, figure \ref{fig:perturbation} shows that the maximum values remains constant, equal to $F(x_0)$, except if $\ell$ becomes so small that the value of $F(x)$ for the minimum value of $k_2\ell$ reaches the same value $0.327$. Using the subscripts $\star$ for referring to the optimal value and $(i)$ ($i=1,2,3$) for referring to the studied criterion, this corresponds to $\ell^{(1)}_\star=0.205\;L\;$. Below this value, the maximum of $F(x)$ grows rapidly.
As a consequence, the optimum of $\IH$ for the criterion 1 is finally the value $\ell^{(1)}_\star=0.205\;L\;.$ Nevertheless, this value is not so critical, because the  criterion function $\ff_1$ for criterion 1 is linear with $\ell$, over the interval $[0.205\;L,L/3]\;,$ thus  the variation is not strong. Moreover, it appears clearly that any other criterion relative to the maximum of the frequency deviation will lead to a value within the interval, because the maximum is reached two times instead of one only.\newline
Finally, taking into account that the interval $[\ell_{min},\ell_{max}]$ corresponds actually to a slightly smaller interval than a twelfth i.e $2^{18/12}$, a numerical study leads to the optimal value for criterion 1 slightly smaller than $\ell=0.205\;L$, i.e $\ell^{(1)}_\star= 0.2041\,L\;.$ Figure \ref{fig:criterion} confirms that for criterion 1, the function $\ff_1$ increases linearly above this optimum value but increases strongly when $\ell$ decreases below it.

The register hole deviation obtained with this ``optimal'' position is plotted on figure \ref{fig:deviation_reg_bis}. As expected, the maximum of deviation appears at the beginning of the register and is almost equal to a quarter tone. Moreover, this result shows that the register hole is located close to its ideal position to produce the \textit{c$\,^\prime$}/\textit{g$\,^{\prime\prime}$} transition: the register hole is located at one third of the effective length for this fingering so that the frequency shift is zero.
\subsubsection{Criterion 2: mean of the square of the derivative of the deviation} 
\begin{figure}[hbtp]
\centering
\includegraphics[width=6cm]{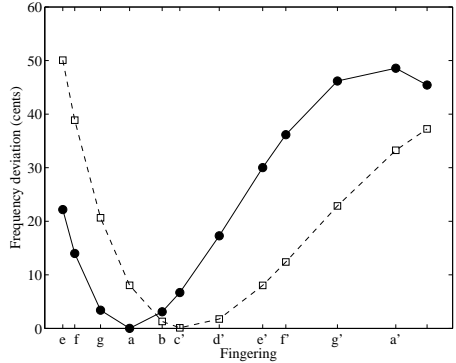}
\caption{\textit{Frequency deviations associated with the register hole. The register hole is found to be located at $\ell^{(1)}_\star=120.45\;mm$ from the top end for criterion 1 ($\,\square\,$) and $\ell^{(2)}_\star=146.2\;mm$ for criterion 2 ($\bullet$).}}
\label{fig:deviation_reg_bis}
\end{figure}

The register hole location achieved with the minimization of (\ref{eq:cout_reg}) is found to be \begin{equation}\frac{\ell^{(2)}_\star}{L}=0.250\;,
\label{eq:res_reg_criterion2}
\end{equation} which is not only larger than the previous result but also larger than the one found with numerical data given by Nederveen ($\ell/L=0.230$). For the investigated clarinet (see table \ref{tab:data}), this ratio gives $\ell/L=0.240$. The frequency deviations obtained with the optimal position defined by (\ref{eq:res_reg_criterion2}) and the one found in literature \cite{Dalmont:95,Benade:76} are very similar: the \textit{a}-\textit{e$\,^{\prime\prime}$} transition is correct and the frequency shift at the beginning of the register is about 20 cents (see figure \ref{fig:deviation_reg_bis}).

\subsubsection{Conclusion}
\begin{figure}[htpb]
\centering
\includegraphics[width=6cm]{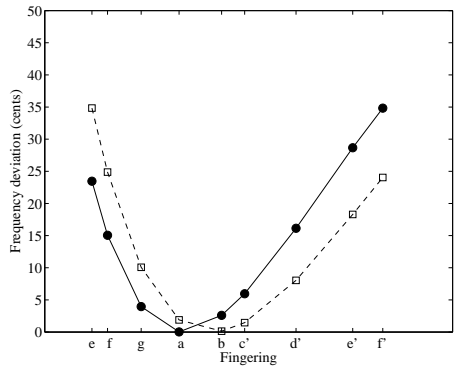}
\caption{\textit{Frequency deviation associated with the register hole in the case of a real clarinet (i.e restricting our optimization to fingering from \textit{e} to \textit{f$\,^\prime$}: above \textit{f$\,^\prime$}, fingerings are modified to ensure the well tuning and the third resonance is used). Optimal positions are $\ell^{(1)}_\star=133.7\;mm$ ($\,\square\,$) and $\ell^{(2)}_\star=145.0\;mm$ ($\bullet$)}}\label{fig:optim_clarinette}
\end{figure}

Performing optimization by minimizing the criterion functions defined by (\ref{eq:cout_reg_max}) and (\ref{eq:cout_reg}) gives different results and indicates that the final location of the register hole is the result of a compromise. However, as expected with equation (\ref{eq:IH_trou_bis}), the magnitude of the maximum of deviation increases with the distance $\ell$ but very slightly for location larger than $\ell/L=0.2041$. As a consequence, above this critical value, the location of the register hole is not so essential on this point of view and this can explained the difference in the register hole location observed between Nederveen's and the investigated clarinets for instance. Finally, it appears that a register hole location far from the effective input and above the critical value should be an interesting compromise in order to accurate the first twelfths of the register. Hence, we choose the location given by the optimization of criterion 2 that is $\ell^{(2)}_\star=146.2\;mm$.\newline
Restricting now attention to fingerings from \textit{e} to \textit{f$\,^\prime$} (and not up to \textit{a$\,^\prime$\#}) may be interesting in order to understand what instrument makers do. The result is that the optimization process converges to a position for the register hole between the two previous extreme values for the two criterion functions. Moreover, for the criterion 1, the distance from the clarinet effective input is found to be $\ell^{(1)}_\star=133.7\;mm$. This result corresponds very well to the location used by Nederveen in its calculation \cite{Nederveen:69} (i.e $\ell=135\;mm$). For that case, the twelfths at the bottom of the scale are still very large but the maximum of the frequency deviation, which is also obtained for the lowest tone, has noticeably fallen to 35 cents (see figure \ref{fig:optim_clarinette}). Concerning criterion 2, the location is very close from the one found for the large compass.

\subsection{Adjustements of natural frequencies by means of small changes of the bore}
\subsubsection{Overview of the possibilities}
As mentioned earlier, the criterion is now to compensate for the frequency deviation due to the register hole by means of small changes of the bore. We are looking for a solution localized in the upper part of the instrument, i.e a solution acting for all fingerings. Looking at table \ref{tab:recapitulatif}, only three systems give inharmonicity in the right  direction: an abrupt change in cross section area with $S^\prime<S$, a change of conicity at the top end, and a localized enlargement or contraction. Concerning the case of a closed cavity, it has been shown (see figure \ref{fig:inharmo_cav_th}) that both positive and negative inharmonicity can be generated: in order to produce negative inharmonicity, the condition $k_1\ell>\pi/4$ which corresponds to playing frequencies larger than $f_1=c/8\ell_{max}$ must be valid. Thus, the accuracy of the twelfths at the end of the second-register scale (from \textit{e'} fingering) would be improved only.

\subsubsection{Abrupt change in cross section area: $S^\prime<S$}
\begin{figure}[h!tbp]
\centering
\includegraphics[width=6cm]{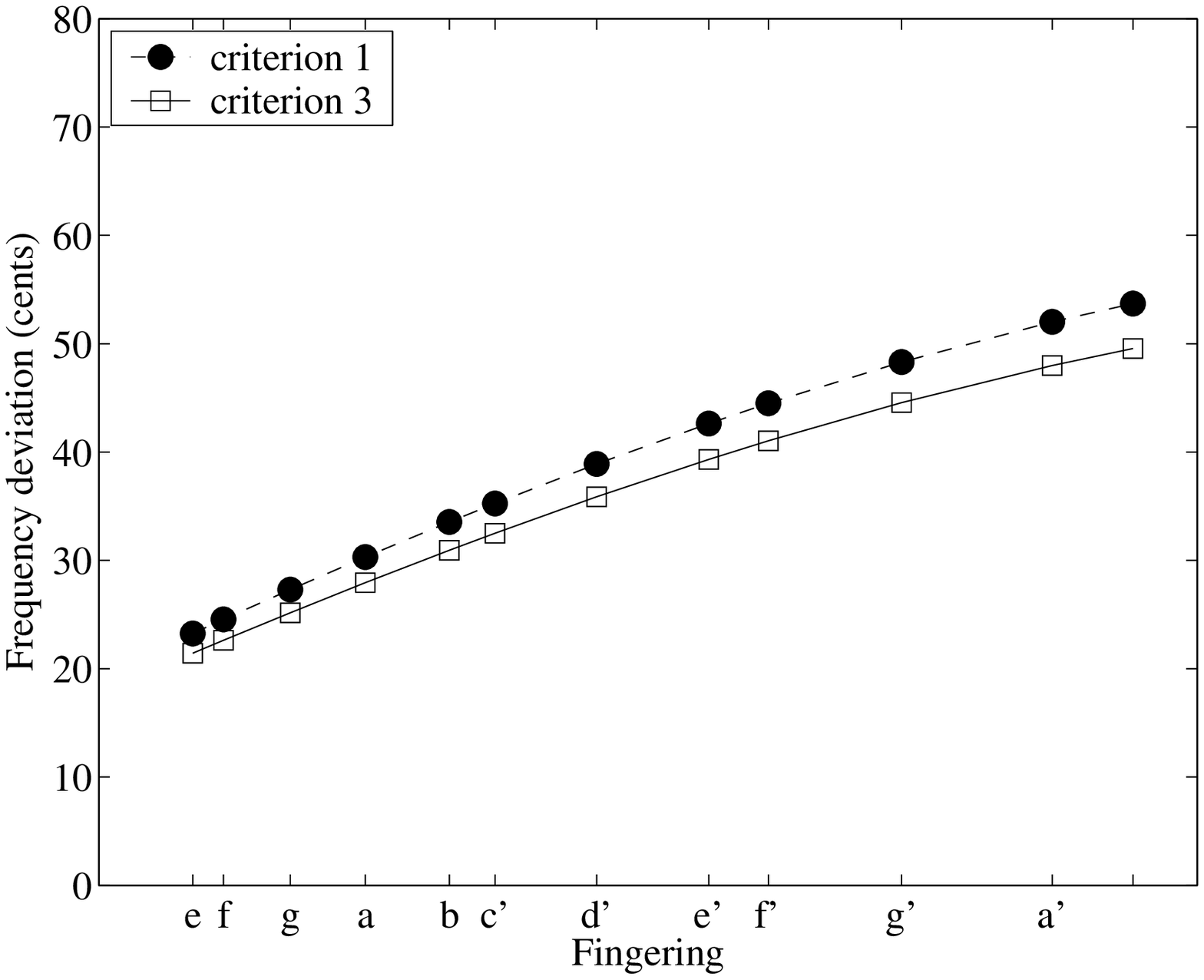} \includegraphics[width=6cm]{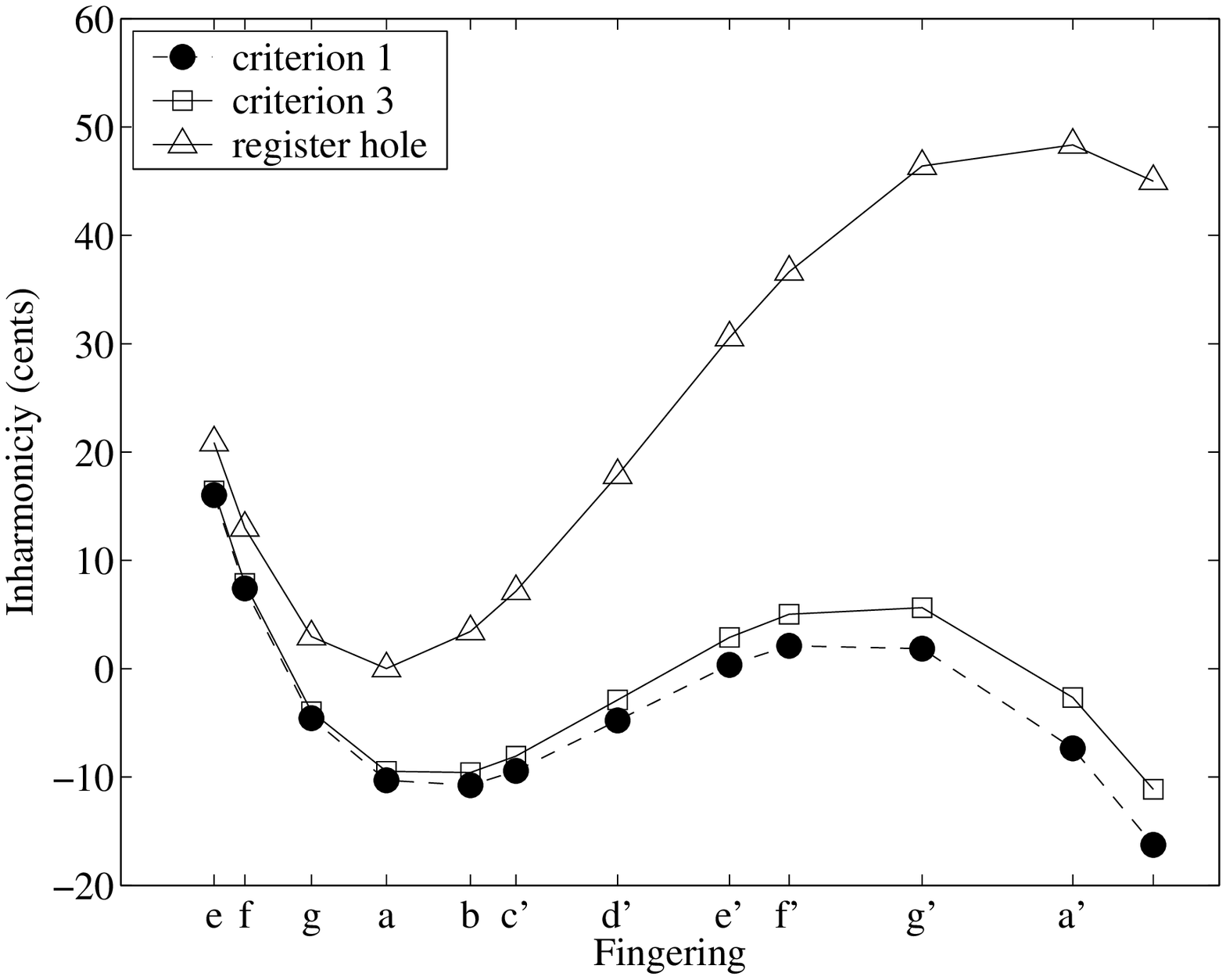}
\caption{\textit{Frequency deviation for the first register (left) and inter-register inharmonicity (right): abrupt change in cross section area.}} \label{fig:disc} \end{figure}

Fixing the radius of the tube downstream the discontinuity equal to $7.5\;mm$, the upstream tube radius $R^\prime$ and the location $\ell$ of the discontinuity are used as optimization variables. Performing optimization leads to the following results : 
\begin{itemize}
\item criterion 1 : $\ell^{(1)}_\star=76.9\;mm$ and $R^{\prime}{}^{(1)}_\star=7.1\;mm$.
\item criterion 3 : $\ell^{(3)}_\star=76.4\;mm$ and $R^{\prime}{}^{(3)}_\star=7.1\;mm$. 

\end{itemize} 
The deviation from the tempered scale of the first resonance frequencies and inter-register for each fingering are plotted on figure \ref{fig:disc}. Results show that even if "optimal" parameters are quite different, the shape and the magnitude of final inharmonicities obtained with the two criterion functions are very similar. The $f_2/f_1$ frequency ratio is noticeably improved for all fingerings except for the twelfth associated with the lowest tone which is still 20-cent large. An interesting result is that the discontinuity in the diameter is found to be located near the barrel joint which is at a distance of $60\pm 4\;mm$ from the closed-end on a clarinet. This result agrees with the practice of many makers: Nederveen noticed that change in cross section area are mostly found  near the embouchure for reed instruments.

\subsubsection{Localized enlargement/contraction}\label{par:elarcont_loc}
\begin{figure}[h!tbp] \centering
\includegraphics[width=6cm]{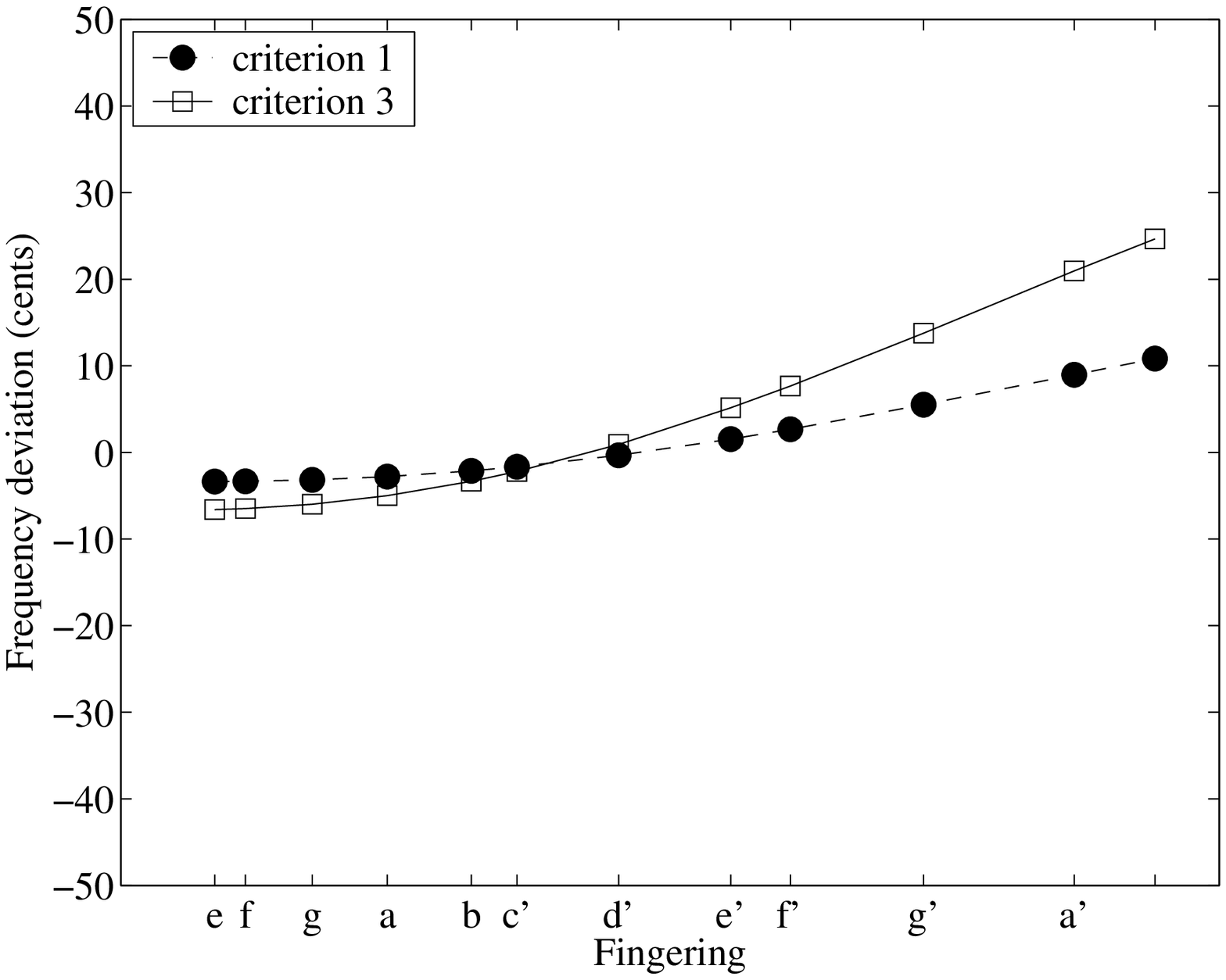}
\includegraphics[width=6cm]{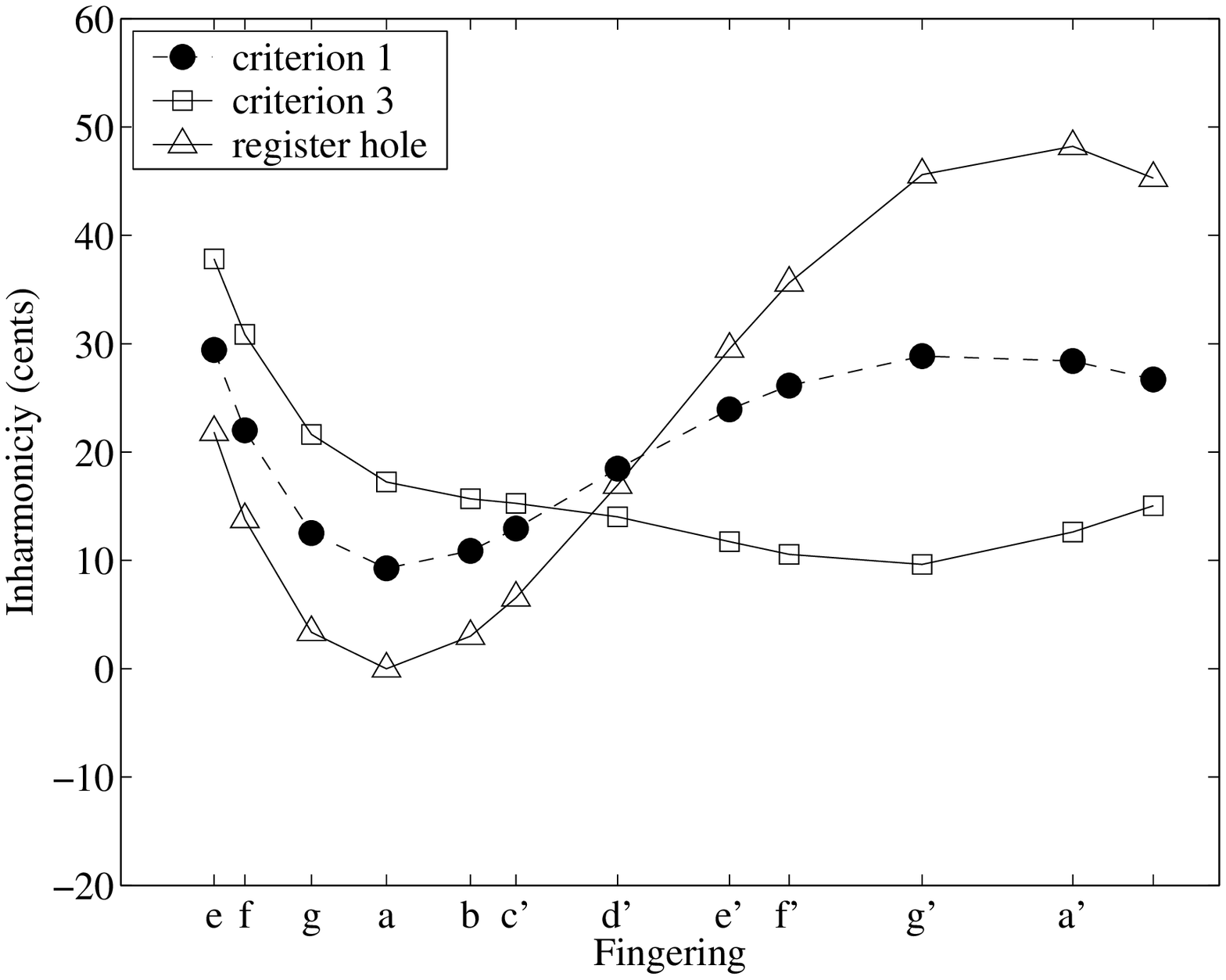}
\caption{\textit{Frequency deviation for the first register (left) and inter-register inharmonicity (right): localized enlargement.}} \label{fig:disc_elar_elarg} 
\end{figure}

\begin{figure}[h!tbp] \centering
\includegraphics[width=6cm]{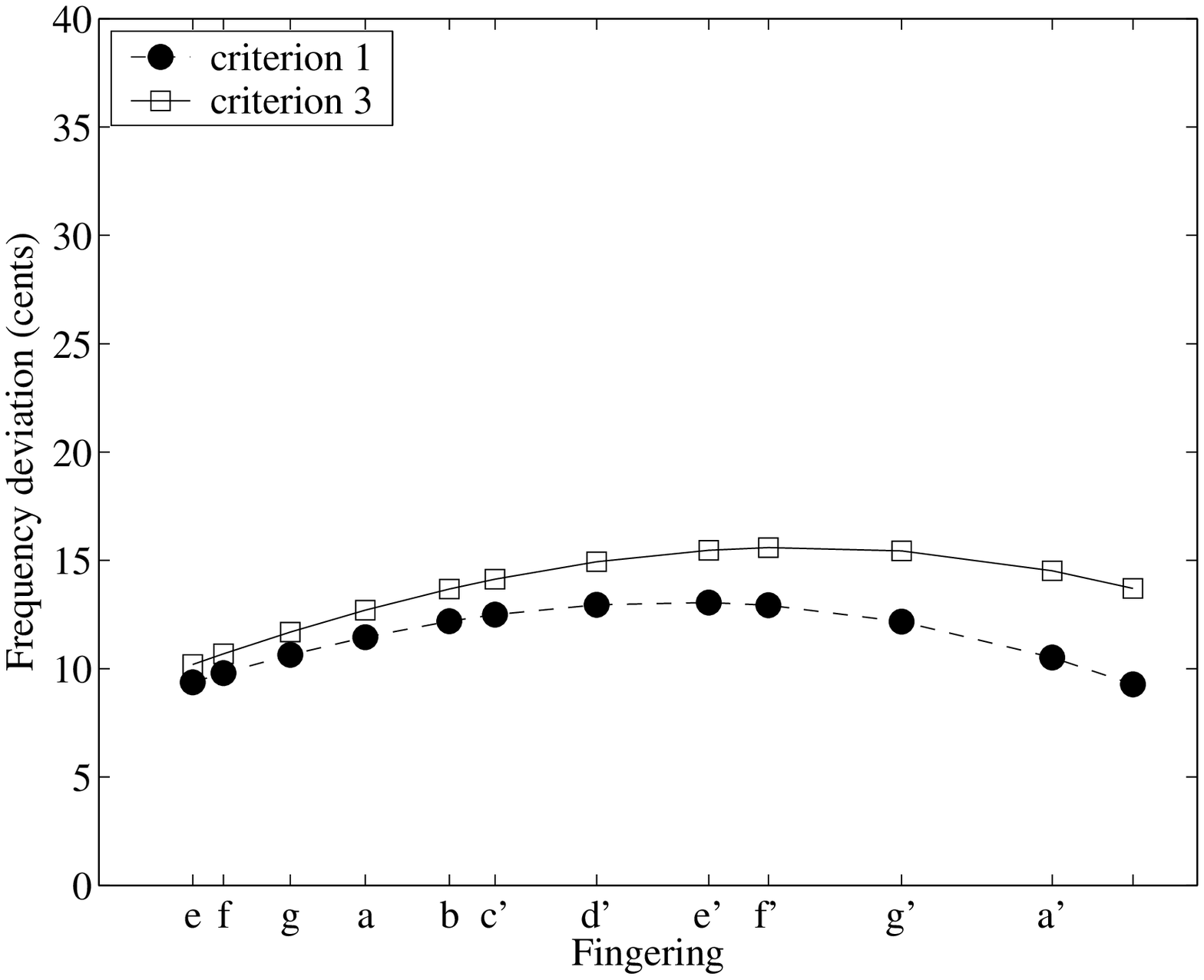}
\includegraphics[width=6cm]{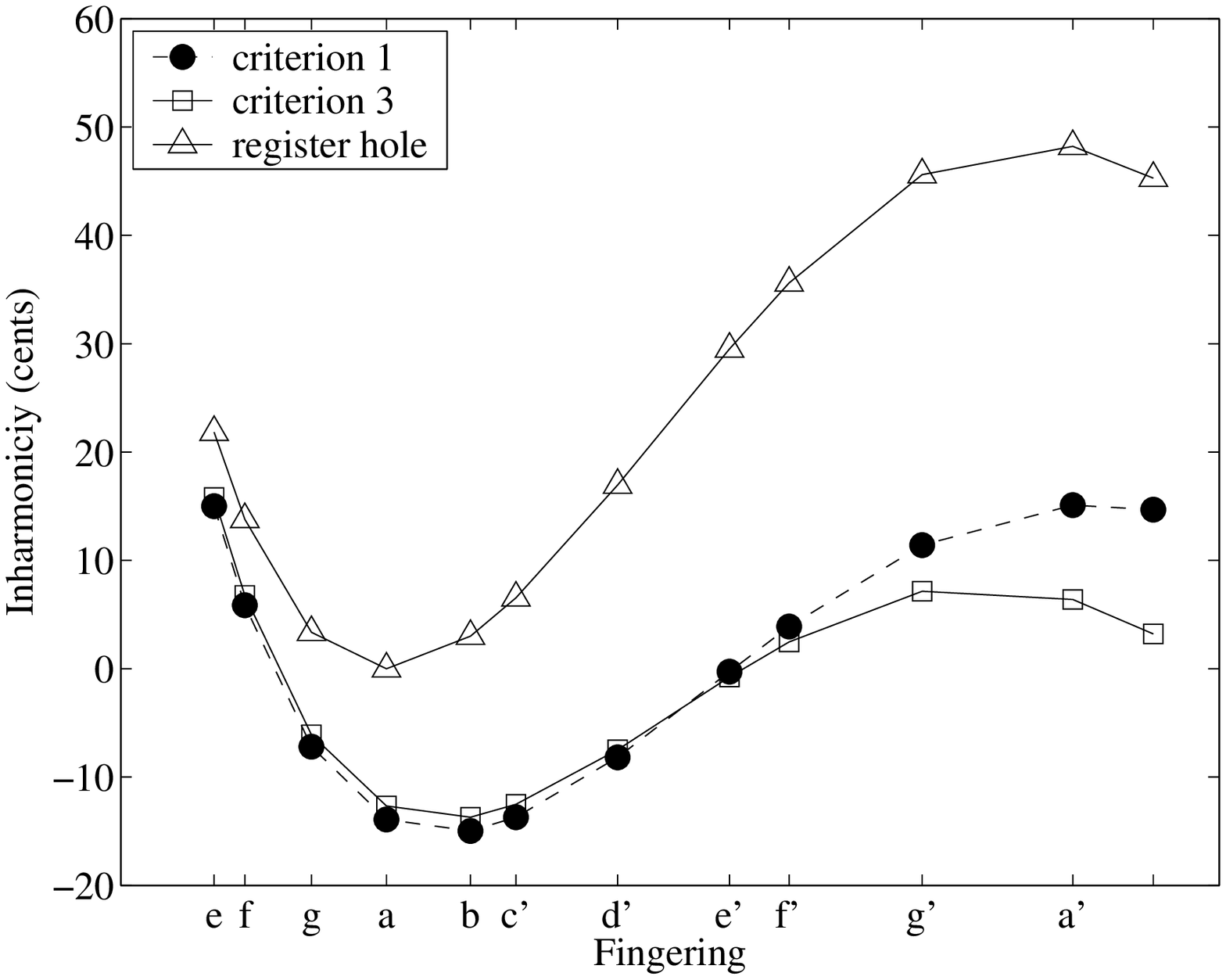} 
\caption{\textit{Frequency deviation for the first register (left) and inter-register inharmonicity (right): localized contraction.}} \label{fig:disc_loc_con} \end{figure}

The optimization variables used in this case are the location $\ell$ of the perturbation, its length $\ell^{\prime}$ and the radius $R^{\prime}$ as indicated on table \ref{tab:deltaL}. Performing optimization for the case of a localized enlargement leads to the conclusion that small enlargement does not improve the relationship between first and second resonance frequencies. Even if optimization process converges with both criterion functions, final inharmonicities are not satisfactory. Figure \ref{fig:disc_elar_elarg} shows the tendency of the bottom notes to widen the $f_2/f_1$ ratio which is musically unacceptable.
On the contrary, for the considered particular case, contracting the air column causes negative inharmonicity across the entire register and tends to improve the inharmonicity (see figure \ref{fig:disc_loc_con}). The optimization process converges  to the following parameters :
\begin{center}
\begin{tabular}{cc} 
& \\
$\mbox{criterion 1} 
\left\{ \begin{array}{l}
\ell^{(1)}_\star=62.74\;mm\\
\ell^{\prime}{}^{(1)}_\star=36.76\;mm\\
R^{\prime}{}^{(1)}_\star=7.17\;mm
\end{array}
\right.
$
&
$\mbox{criterion 3} 
\left\{\begin{array}{l}
\ell^{(3)}_\star=60.0\;mm\\
\ell^{\prime}{}^{(3)}_\star=22.41\;mm\\
R^{\prime}{}^{(3)}_\star=7.0\;mm
\end{array}
\right.
$\\
 & \\
\end{tabular}
\end{center}
When minimizing the criterion function defined by equation (\ref{eq:cout_opt_max}), an interesting result is that the design parameters given by the optimization process are very similar to those of a clarinet barrel. 
As for an abrupt change in cross section, the contraction is located at a point near the barrel joint. In addition, the acoustic length of the contraction is the same as a common barrel ($\ell_{barrel}\sim40\;mm$). When looking at what makers do, their practice varies considerably concerning the geometry of the barrel. However, some makers insert narrower barrel than the cylindrical portion of the entire instrument in order to get accurate twelfths in the upper part of the scale \cite{Brymer:76}.
While the results given by the minimization of (\ref{eq:critere}) generate negative inharmonicity (see figure \ref{fig:disc_loc_con}), the optimization process converges to two lower boundaries ($R^{\prime}{}^{(3)}_\star=R^{\prime}_{max}$ and $\ell^{(3)}_\star=\ell_{min}$). When enlarging the domain where optimization variables are looking for, a new minimum (lower than the one found) is achieved. Thus, we do not consider this solution in the next. Finally, the mean value of inharmonicity is about -0.6 cents and the standard deviation of the mean value is equal to 11.9 cents.

\subsubsection{Change in conicity at the input}  \label{par:insertion_cone}
\begin{figure}[htbp] \centering
\includegraphics[width=6cm]{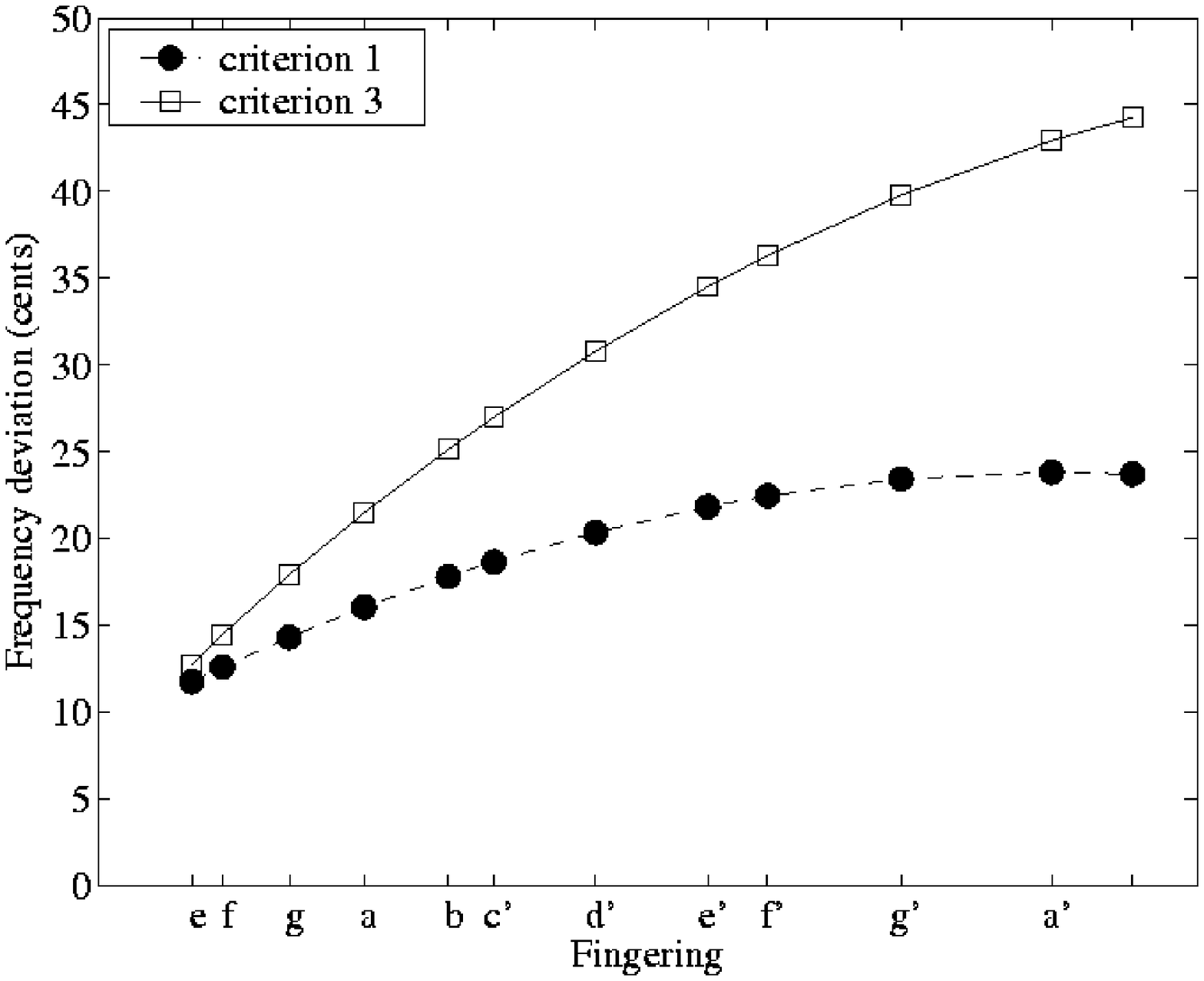}
\includegraphics[width=6cm]{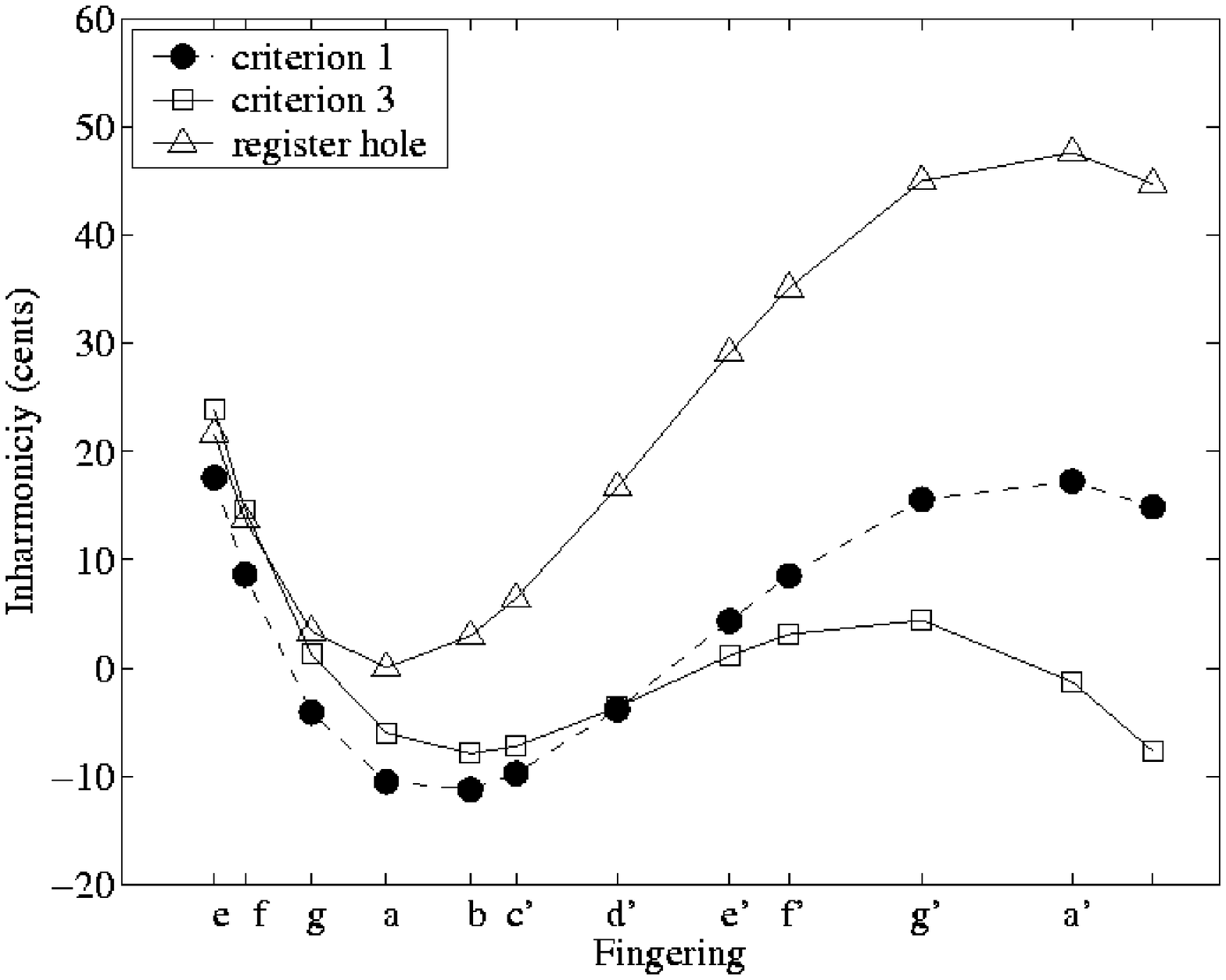}
\caption{\textit{Frequency deviation for the first register (left) and inter-register inharmonicity (right): positive truncated cone.}} \label{fig:reg_trois_cone} \end{figure}

The optimization variables used for the case of tapered perturbation (see table \ref{tab:deltaL}) are the radius of the upper end of our model $R^{\prime}$, the top angle $\theta$ and the location point $\ell$.
For each criterion function, the optimization process converges to the following design paremeters :
\begin{center}
\begin{tabular}{cc} 
& \\

$\mbox{criterion 1} 
\left\{ 
\begin{array}{l}
\ell^{(1)}_\star=98.5\;mm\\
\theta^{(1)}_\star=0.10^\circ\\
R^{\prime}{}^{(1)}_\star=7.46\;mm
\end{array}
\right.
$ 
& 
$\mbox{criterion 3} 
\left\{ 
\begin{array}{l}
\ell^{(3)}_\star=73.05\;mm\\
\theta_\star^{(3)}=0.22^\circ\\
R^{\prime}{}^{(3)}_\star=7.37\;mm\\
\end{array}
\right.
$\\
 & \\
\end{tabular}
\end{center}
The lengths of the tapered perturbations corresponding to criterion 3 and criterion 1 are about 44.5 mm and 46.5 mm which are both larger than the length of a classic barrel. Concerning the location point, it is near the barrel joint for criterion 3 and slightly below this point for criterion 1. An important thing to note is that the location achieved by the optimization process is always close to the barrel joint whatever small change is used. \newline
Figure \ref{fig:reg_trois_cone} shows final inharmonicities obtained with the minimization of the two criterion functions. The $f_2/f_1$ frequency ratio is improved globally  across the entire scale for both criterion functions except for the first fingering which is deteriorated slightly for criterion 3. As a consequence, only the design parameters given by the minimization of criterion 1  are considered. The mean value of the total inharmonicity is found to be 3.6 cents and the standard deviation of the mean value is about 11 cents. \newline
Finally, results indicate that the 20-cent deficiency still remains for the notes at the bottom of the scale whatever acoustic system is used in order to alter the resonance frequencies. Thus, no improvement can be done to compensate for the effect of the register hole except for tones at the end of the scale. 

\section{Conclusion}
Many deviations from the standard cylindrical tube can be observed on a woodwind instrument. The present work provides some theoretical results concerning the effects of these bore irregularities  in the perspective of the relationship between the first two resonance frequencies for the two first registers. The calculations using small perturbations are classical but the case of a side hole has been treated precisely, taking into account both the series and shunt impedances.\newline
Calculations of inharmonicity associated with small perturbations and application to a real case give access to the relative influence on the first two resonance frequencies of each pertubating system found on a woodwind and reveal, in the case of the clarinet, the necessity to resort to bore perturbations to compensate for the damage on the resonance frequencies caused mainly by the presence of closed holes and the register hole.\newline
In addition, the use of simple optimization techniques has been shown to be useful in order to give a more general view of the problem. However, optimization is very sensitive to the definitions of the criteria; for our purpose, performing optimization on the accuracy of the twelfths does not impose a unique possible criterion. For instance, the use of optimization techniques with weight for the lowest tones is of course possible and would probably give other results.\newline
The main results of the present work can be summarized as follows:
\begin{description}
\item - the origin of the occurence of wide twelfths for the lowest tones of a clarinet is not to be found in shifts of the resonance frequencies of the resonator, as opposed to common wisdom. A way for future investigations may be the influences of the blowing parameters, reed parameters and the vocal tract since players can modify both of them and claim to do it.
\item - with a single corrective system, designing a new instrument having for the two registers a compass  of a twelfth with the same fingering except the register hole opening, is impossible without tuning deficiencies;
\item - if the aim is the improvement of a common clarinet, the use of contractions near the barrel can help to some extent; nevertheless, no simple solution located in the upper part of the instrument can ensure the perfect tuning of the twelfths corresponding to the first lowest tones.
\end{description}
Finally, combining global and local solutions, i.e solutions acting mainly on one particular tone, is necessary in order to achieve the best tuning.

\section*{Acknowledgements}
We wish to thank P. Mallet, J.P. Dalmont and O. Gipouloux for fruitful discussions, as well as J.P. Dalmont for a loan of the measurement device. The Laboratoire Kastler Brossel of the D\'epartement de Physique de l'ENS is a laboratoire associ\'e au CNRS et \`a l'Universit\'e Pierre et Marie Curie.
\section*{Appendix}
\appendix
\numberwithin{equation}{section}
\section{Exact calculation of length corrections}\label{anx:un}
The aim of the appendix \ref{anx:un} is to give some precise definitions and exact formulae for some elements often encountered on wind instruments. Exact formulae can be useful for the computation of resonance frequencies by using an iterative procedure. The geometries of the different elements are shown in table \ref{tab:deltaL}.

\subsubsection*{Abrupt change in cross section}
\noindent
Looking for a equivalent system of section $S$, one can write at the discontinuity
$$-j Y_{c}^\prime\tan k\ell=-j Y_{c}\tan k(\ell+\Delta \ell)$$ where $Y_{c}^\prime=S^\prime/\rho c$ and $Y_{c}=S/\rho c$. Manipulating the above expression gives result in 
\begin{equation} k\Delta \ell=
\arctan \left(\d\frac{\frac{1}{2}(\alpha-1)\sin 2k\ell}{1+(\alpha-1)\sin^2 k\ell}\right)\;,
\label{eq:deltaL_disc_exact}\end{equation} where $\alpha=S^\prime/S$ is the ratio of cross sections of the tubes located upstream and downstream the discontinuity.

\subsubsection*{Localized enlargement/contraction at the input of the instrument} 
\noindent
Considering a localized enlargement or contraction of length $\ell^\prime$ and cross section area $S^\prime$ located at point $\ell$ and denoting $\alpha=S^\prime/S$  the change in diameter where $S$ is the cross section area of the main tube, it can be written: 
$$\left\{ 
\begin{array}{l}
Y_{up_1}=-jY_c \tan k\ell\quad \hbox{at point $\ell$}\\
Y_{down_2}=-jY_{c} \tan k(\ell+\ell^\prime+\Delta \ell)\quad\hbox{at point $\ell+\ell^\prime$}\;.\\
\end{array}
\right.$$
which results in after some algebra 
$$
k\Delta \ell=\arctan\left(\tan k\ell^\prime\;\frac{T T_2-\alpha(1+T T_2)+\alpha^2}
{-T\tan k\ell^\prime+\alpha(1+T T_2)+\alpha^2\tan k\ell^\prime T_2}\right)\;,
$$
where $T= \tan k\ell$ and $T_2=\tan k(\ell+\ell')$. \newline
Denoting $\alpha=1+X$, the expression of the length correction becomes
\begin{equation}
k\Delta \ell=\arctan \left(X \sin k\ell^{\prime}\frac{\cos k(2\ell+\ell^{\prime})+X \cos k\ell \cos k(\ell+\ell^\prime)}
{1+X \cos^2 k\ell^{\prime} + X (X+2)\sin k\ell^\prime \cos k\ell \sin k(\ell+\ell^\prime)})\right.\;.
\label{eq:deltaL_localized_exact}
\end{equation}
In the limit that $\alpha$ is close to the unity, an approximation of equation (\ref{eq:deltaL_localized_exact}) is given by equation (\ref{eq:deltaL_disc_loc}).

\subsubsection*{Truncated cone at the input}
\noindent
This paragraph deals with the case of a perturbation in taper of length $\ell_c$ located at distance $\ell$ from the reed tip. Calculations for a positive and a negative taper are identical. The unique difference is that the sign of both quantities $X$ and $X^\prime$ has to be inverted.\newline
Assuming the quantities $x_i$ to be positive and $x_2>x_1$, and noting that the two inductances located at points $\ell$ and $\ell+\ell_{c}$ are positive and negative respectively, the following system is derived :$$\left\{ \begin{array}{l} Y_{up_2}=Y_{2}+Y_{down2}\quad\hbox{at
point $\ell+\ell_c$ ($x_2>0$) and $L>0$}\;,\\ Y_{up_1}=-Y_{1}+Y_{down}\quad \hbox{at point $\ell$
($x_1>0$) and $L<0$}\;,\\ Y_{down1}=-jY_c \tan k(\ell+\ell_{c}+\Delta \ell)\;, \end{array}\right.$$ where
$\Delta \ell$ is the global length correction.\newline
Denoting $X=\d\frac{\ell}{x_1}$ and $X^\prime=\d\frac{\ell+\ell_c}{x_2}$, the result is
\begin{equation}
k\Delta \ell=\arctan\,\Lambda\;, \label{eq:deltaL_exact_dv}\end{equation} where $$\Lambda=\left(
\d\frac{X\frac{\cos^2 k(\ell+\ell_c)}{k(\ell+\ell_c)}-X^\prime\frac{\cos^2
k\ell}{k\ell}+XX^\prime\frac{\cos k\ell}{k\ell}\frac{\cos k(\ell+\ell_c)}{k(\ell+\ell_c)}
\sin k\ell_c}{1-\frac{1}{2}X^\prime\frac{\sin 2k\ell}{k\ell}+\frac{1}{2}X\frac{\sin
2k(\ell+\ell_c)}{k(\ell+\ell_c)}+XX^\prime\frac{\cos k\ell}{k\ell}\frac{\sin k(\ell+\ell_c)}{k(\ell+\ell_c)}\tan k(\ell+\ell_c)\sin k\ell_c}\right)\,.$$
 In the limit of small perturbation, the length correction for a truncated cone is the sum of the length correction associated with both changes in taper.
\numberwithin{equation}{section}
\section{Effect of the resistive term in the hole impedance on resonance frequencies}\label{anx:deux}
The aim of this appendix is to give an analytic expression of the playing frequency in term of the resistive part $R$ of the hole impedance. This can be interesting especially when resistive effects are large and can occur at high level when nonlinear effects appear, especially for narrow  holes, like the register hole. This appendix describes the calculations which lead to the following expression for the frequency of the played tone 
\begin{equation}
f=\frac{f_o+A f_c}{1+A}
\label{eq:freq_anal}
\end{equation}
where $f_o$ and $f_c$ are the frequencies obtained when the resistance term is zero and infinite respectively (when $R$ is infinite, the effect of the hole disappears as it was closed), and $A$ is a dimensionless factor given by $$A=\left(\d\frac{2R}{\pi}\right)^2\d\frac{1}{\ell_{hole}}
\d\frac{L\ell}{\ell_d+\Delta \ell}\;,$$ with $L$ the total length of the resonator, $\ell$ the hole location, $\ell_{hole}=h^{\prime}S/S_h$ and $\Delta \ell$ the equivalent length of the hole. As done in this study for the case of open hole, the series impedances $Z_a$ are assumed to be small (see section \ref{subsub:sidehole}).

\begin{figure}[htbp] 
\centering
\includegraphics[width=8cm]{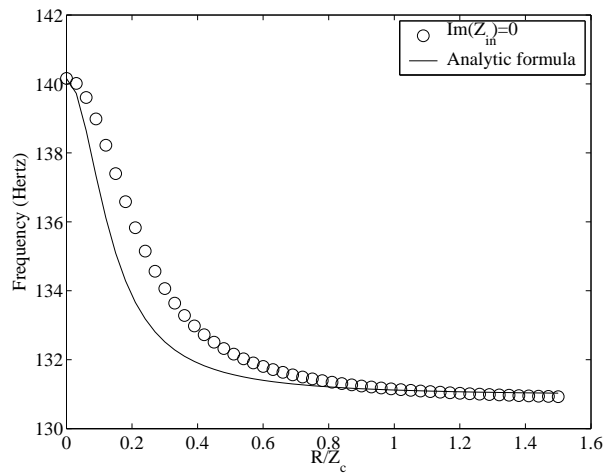} 
\caption{\textit{Playing frequency as a function of the real part of the tone hole impedance ($R=10\;mm$, $r=7\;mm$,$h=9\;mm$). In the limit that $R\rightarrow 0$, the frequency of the played tone tends to $f_o$,
the frequency obtained when the hole is open. In the limit that $R\rightarrow \infty$, the frequency of the played tone approaches $f_c$, the frequency when the hole being closed.}} 
\label{fig:ana_freq} 
\end{figure}

\paragraph*{}Using dimensionless impedance quantities, the impedance equivalent $Z_{eq}$ to the tone hole impedance and to the tube downstream the discontinuity can be written at the discontinuity point as
$$Z_{eq}=Z_h\,//\,Z_{down}\;,$$ where $Z_h=R+jX$ is the tone hole impedance and $Z_{down}$ the impedance of the tube donwstream the discontinuity. Then, the input impedance $Z_{in}$ of the system can be derived  as follows:
\begin{equation}
Z_{in}=\frac{-t+jZ_h T\left(\d\frac{1+t^2}{T-t}\right)}{Z_h
\left(
\d\frac{1+t^2}{T-t}\right)+j}\:,
\label{eq:imped}
\end{equation}
where \begin{itemize}
\item   $Z_{in}$ is the input impedance of the system\:;
\item   $Z_h=R+jX$ is the tone hole impedance\:;
\item   $t=\tan k\ell$\:;
\item   $T=\tan kL$\:.
\end{itemize}
For the resonance frequencies, the imaginary part of the input impedance vanishes, i.e
\begin{equation}
\mathcal{I}m(Z_{in})=0 \Longleftrightarrow
\left(\frac{T-t}{1+t^2}+X\right)\left(t\frac{T-t}{1+t^2}+XT\right)+TR^2=0\;.
\label{eq:imag}
\end{equation}
Denoting
$$A=\sin k(L-2\ell)+\sin kL+2\chi \cos kL\;,$$
$$B=\cos k(L-2\ell)-\cos kL+2\chi \sin kL\;,$$
$$C=2R^2\sin 2kL\;,$$
where $\chi=k\ell_{hole}$, equation (\ref{eq:imag}) can be rewritten in the form \begin{equation}
AB+C=0\;.\newline
\label{eq:AB}
\end{equation} 
By writing $kL=k_1L+\epsilon_1$ where $k_1$ is the wavenumber of the fundamental frequency of the played tone defined as $A=0$ (when $R$ vanishes):
\begin{equation}
\sin k_1(L-2\ell)+\sin k_1 L+2 \chi \cos k_1 L=0\;,
\label{eq:annul}
\end{equation}
the use of Taylor's formula to the first order for the A and B quantities results in:
\begin{multline}
A=\epsilon_1[(L-2\ell)\cos k_1(L-2\ell)+L\cos k_1 L+2\ell_{hole}\cos k_1L
-2\chi L\sin k_1L]\;,
\end{multline}

\begin{multline}
B=\cos k_1(L-2\ell)-\cos k_1L+2\chi\sin k_1L\\
+\epsilon_1[L\sin k_1L-(L-2\ell)\sin k_1(L-2\ell)
+2\ell_{hole}\sin k_1L\\ +2\chi L\cos k_1L] \;.
\end{multline}
Neglecting terms of second order in $\epsilon_1$ and using equation (\ref{eq:annul}), the quantity $AB$ becomes:
\begin{multline}
AB=\left(\d\frac{1}{\cos k_1\ell_{hole}}[L(1+\cos 2k_1\ell)+2\ell_{hole} \cos^2 kL-2\ell\cos k(L-2\ell)  \cos kL\right)  \\
\left(\d\frac{\cos 2k_1\ell_{d}-1}{\cos k_1L}\right).
\end{multline}
In the limit of $k_1\ell_{d}\ll1$ (i.e $\d\frac{\ell_{d}}{\ell}$ and $\d\frac{\Delta \ell}{\ell}$ are smaller than unity) and noting $k_1=\d\frac{\pi}{2(\ell+\Delta \ell)}$, we have:
\begin{eqnarray}
&&\cos k_1L=-\d
\frac{\pi}{2}\d\frac{\ell_{d}{^2}}{\ell(\ell_{hole}+\ell)}+o(\d
\frac{\ell_{d}}{\ell}) \nonumber\\
&&\cos 2k_1\ell_d=1+o(\d\frac{\ell_{d}}{\ell}) \nonumber\\
&&\cos k_1(L-2\ell)=\sin \d
\frac{\pi}{2}\d\frac{\Delta \ell+\ell_{d}}{2}=\d
\frac{\pi}{2}\d\frac{\Delta \ell+\ell_{d}}{2}+o(\d \frac{\ell_{d}}{\ell}) \nonumber\\
&&\cos 2k_1\ell=-1+o(\d \frac{\ell_{d}}{\ell}) \;.\nonumber
\end{eqnarray}
Substituting these results in equation (\ref{eq:AB}) leads to the following equation
$$\left(\pi^2 \ell_{hole} \d\frac{\ell_d+\Delta \ell}{\ell}\right)\epsilon_1+2 R^2 \sin
2kL=0\;.$$
Noting $kL=k_cL+\epsilon_2\quad$ where $\quad k_cL=\d\frac{\pi}{2}$ ($k_c$ is the wavenumber when the hole is closed), it can be written :
\begin{equation}
\pi^2 \ell_{hole} \d\frac{\ell_d+\Delta \ell}{\ell}(k-k_1)+4 R^2 L(k-k_c)=0\;,
\end{equation}
therefore,\begin{equation}
f= \frac{f_o+A f_c}{1+A}\;,
\label{eq:freq_anal2}
\end{equation}
where \begin{equation}
A=\left(\d\frac{2R}{\pi}\right)^2\d\frac{1}{\ell_{hole}}
\d\frac{L\ell}{\ell_d+\Delta \ell}\:,\end{equation}
$f_o$ is the frequency of the tone with no resistance, and $f_c$ is the frequency of the tone with an infinite resistance term.\newline
A careful limiting process shows that all is in order in the two extreme cases i.e when $R\rightarrow 0$ and $R\rightarrow\infty$. In the limit that $R$ goes to zero, equation (\ref{eq:freq_anal2}) gives the frequency to tend to $f_o$, the frequency of the tone when the hole is open. In the opposite limit when $R$ tends to infinity, equation (\ref{eq:freq_anal}) gives the frequency to be $f_c$. Figure \ref{fig:ana_freq} shows that the playing frequencies given by equation (\ref{eq:freq_anal2}) coincide quite well with the playing frequencies obtained by the zero values of the imaginary part of the input admittance.
Moreover, in the limit that $R$ is much smaller than unity, it follows
\begin{equation}
f=f_o+A(f_f-f_o)\;.
\label{eq:fonction}
\end{equation}
With the use of experimental values obtained by Dalmont et al. \cite{Dalmont:02} about the non linear behavior of an open side hole, it appears that the resistive term of the shunt impedance has small effects in the determination of the playing frequency. For instance, a 5-cent difference is obtained with $R/Z_c=0$ and $R/Z_c=0.05$.

\bibliographystyle{unsrt}
\bibliography{biblio100304}

\end{document}